# Garnet pyroxenites explain high electrical conductivity in the East African deep lithosphere

Thomas P. Ferrand[1,2,3]

1: Institute of Geophysics & Planetary Physics, Scripps Institution of Oceanography, UC San Diego, La Jolla, USA;

2: Institut des Sciences de la Terre d'Orléans, CNRS UMR 7327, Université d'Orléans, France;

3: Institüt für Geologische Wissenschaften, Freie Universität Berlin, Malteserstraße 74-100, Berlin 12249, Germany.*

*current location. Contact: *thomas.ferrand@fu-berlin.de*

## Abstract

A large contrast in electrical conductivity has been observed within the lithospheric mantle between the Tanzanian craton and the Mozambique belt at depths > 70 km. Such contrasts are typically attributed to changes in volatiles and/or melt content, with changes in mineralogy deemed insufficient to impact conductivity. To test this assumption, electrical conductivity measurements were conducted at pressure-temperature conditions relevant to the Tanzanian lithosphere (1.5 and 3 GPa; from 400 to > 1500°C) on xenoliths from Engorora, Northern Tanzania. I observe that, once garnet becomes stable in fertile mantle rocks (> 60 km, 1.7 GPa), it nucleates at grain boundaries and forms the backbone of a conductive network. At 3 GPa, the presence of garnet-rich networks increases conductivity by a factor of 100 regardless of temperature. Numerical models demonstrate that low (< $10^{-2}$ Sm$^{-1}$) and high (> $10^{-1}$ Sm$^{-1}$) conductivities in the Tanzanian lithospheric mantle are best explained by low and high degrees of garnet connectivity, respectively. Such high electrical conductivities in cratonic roots can be explained by the presence of connected garnet clusters, suggesting mantle refertilization. I suggest that the garnet network represents the result of plume triggered metasomatism due to plume impingement, and may represent an unstable structure that will eventually lead to loss of cratonic lithosphere.

The structure and composition of the cratonic lithosphere has been debated for decades (e.g. Takahashi, 1990; Hirth et al., 2000; Evans et al., 2011) but these deep regions of the Earth are only directly accessible via volcanically-derived xenoliths (e.g. Henjes-Kunst & Altherr, 1992; Rauch & Keppler, 2002; Vauchez et al., 2005; Doucet et al., 2014; Chin, 2018) and through indirect geophysical imaging (e.g. O'Donnell et al., 2013; Selway, 2015; Sarafian et al., 2018). The formation of cratons and their evolution through time are still under investigation, particularly their apparent longevity, although focus is also turning to processes of plume impingement that might eventually trigger craton disruption (Foley, 2008; Pernet-Fisher et al., 2015; Wang et al., 2015; Koptev et al., 2016). Magnetotellurics (MT) offers a powerful approach to understanding the structure of the lithosphere and upper mantle. For example, several studies report anomalously high conductivities in the lithospheric mantle (Selway, 2015; Sarafian et al., 2018), notably below the East African Rift (**Fig.1**; Selway, 2015; Sarafian et al., 2018). These anomalies are thought to provide evidence for the presence of volatiles and/or the occurrence of melting at depth (Selway, 2015). A high-conductivity anomaly (> $10^{-2}$ $Sm^{-1}$) is observed below the Tanzanian Craton at depths > 70 ±10 km, while the conductivity of the neighboring Mozambique Belt at the same depth is much lower (< $10^{-3}$ $Sm^{-1}$; Selway, 2015). The contrast between the cratonic root and the mobile belt is striking, and counter to almost all observations in other similar tectonic settings (e.g. Evans et al., 2011; 2019; Sarafian, 2018). Selway et al. (2014) hypothesized that the high conductivity of the cratonic lithosphere is caused by the presence of water dissolved in olivine. However, this explanation contradicts models of cratonic root formation that consider predominantly depleted and dry roots. Furthermore, volatiles are known to lower mantle viscosity, enhancing the likelihood of mantle deformation (Hirth & Kohlstedt, 2003), which would not be consistent with craton stability. Most xenoliths have revealed that the cratonic lithosphere is typically characterized by at best a moderate $H_2O$ content (Udachnaya: Doucet et al., 2014; Kaapvaal: Peslier et al., 2010; Baptiste et al., 2012). The limited $H_2O$ contents observed in Kaapvaal xenoliths are consistent with observed low conductivities (Evans et al., 2011; Jones et al., 2012), while significantly drier cratons exist (e.g. North China craton; Chin et al., 2020). In addition, pyroxenites from Hawaii, located over a mantle plume, are 4 times more hydrated than typical oceanic mantle (Bizimis & Peslier, 2015), which suggests that plume impingement could also increase the water content of cratonic roots.

To probe the origin of electrically-conductive cratonic environments, I experimentally reproduced the pressure-temperature conditions relevant to the deep Tanzanian lithosphere and performed electrical measurements on xenoliths from the area (***Methods, sections M1-M2***). The xenoliths originate from Engorora, Northern Tanzania (Chin, 2018). I used powders of a clinopyroxenite (ENG7) and a dunite (ENG8), two endmember lithologies in the area (Chin, 2018), hereafter referred to as fertile and depleted compositions, respectively (**Table S1**).

High-pressure, high-temperature experiments were conducted using the 14/8 COMPRES electrical assembly (**Fig.S1**) designed for conductivity measurements in a multi-anvil apparatus (Pommier et al., 2019). Pressures of 1.5 and 3 GPa were targeted in order to monitor conductivity in the spinel and garnet stability fields, respectively, and temperature ranged from ≈ 400 to ≈ 1550°C. A summary of the experiments is provided in **Fig.2** and details are available in ***Methods***, **Fig.S2** to **Fig.S4 Table S2** and **Table S3**.

At 1.5 GPa (≈50 km depth) and over the investigated temperature range, the pyroxenite and dunite samples present similar conductivity values. These values would agree with a moderate bulk hydration (~100 ppm; Wang et al., 2006; Yang et al., 2011; Gardés et al., 2014). At 3 GPa (≈100 km depth), dunite conductivity decreases by a factor ranging from 10 to 100, depending on temperature (**Fig.2b**), whereas pyroxenite conductivity increases by a factor of about 10. The results at 3 GPa are in relatively good agreement with conductivities reported previously for dunite, lherzolite and pyroxenite containing minor amounts of $H_2O$ (~ 100 ppm), showing that pyroxenites are significantly more conductive (Wang et al., 2008), although less conductive than ENG7 (**Fig.2a**). At high temperature (1300°C) the pyroxenite is as conductive as a silicate melt with moderate hydration (0.2 – 1 $Sm^{-1}$). The ENG7 and ENG8 samples are likely moderately hydrous, but it cannot explain the contrast between the measured conductivity from 1.5 to 3 GPa (Dai et al., 2012; Gardés et al., 2014).

The presence of garnet is seen in microstructural images of samples retrieved after experiments at 3 GPa (**Fig.3**). Garnet, which represents some vol.% at 3 GPa (up to ~5 vol.% in **Fig.3b**), is preferentially distributed along grain boundaries, where reactive phases are in contact (***Suppl. text***, ***section S1b***) and where diffusivity is enhanced (Fisher, 1951). In the field, garnet networks have been observed in xenoliths (Henjes-Kunst & Altherr, 1992) in the same region (Chyulu, Southern Kenya; **Fig.1**), suggesting that the experiments at 3 GPa would reproduce a natural texture and its bulk electrical response.

The electrical data highlight a transition in the bulk electrical response between 1.5 and 3 GPa, characterized either by an increase (pyroxenite) or a decrease (dunite) in conductivity. Both xenoliths likely contain minor graphitic carbon impurities located at grain boundaries (Watson, 1986; Pearson et al., 1994), which would be consistent with $CO_2$-rich magmatism observed in the area and oxygen fugacity considerations (***Suppl. text, section S2***). Graphite is known to be highly conductive (~$10^5$ $S.m^{-1}$; Duba, & Shankland, 1982) and thus significantly enhance conductivity when connected (≥ 1 wt.%; Wang et al., 2013). However, a recent study revealed that grain-boundary films are not stable and that graphite is thus unlikely to explain the high-conductivity anomalies revealed by MT surveys in the upper mantle (Zhang & Yoshino, 2017). In addition, there is no reason for graphite to become more conductive/connected with increasing pressure (**Fig.2**). As a consequence, the presence of graphite alone is insufficient to explain the data. Instead, microstructural observations illustrate that this electrical transition correlates with the nucleation of garnet, which resides preferentially in the grain-boundary network of the pyroxenite samples (**Fig.3**). In the dunite samples, garnets are scarce and isolated (**Fig.S8**). Some isolated grains of accessory phases are also observed in the samples (e.g. chromite, hibonite, metal alloys) but the spinel-garnet transition (≈ 1.8 GPa) is the major metamorphic transformation affecting mantle rocks at these conditions (review, Ferrand, 2019) and therefore best explains the measured electrical transition between 1.5 and 3 GPa.

Electrical conductivity data suggest that the garnet-rich network observed at 3 GPa in the pyroxenite is highly conductive (> $10^{-2}$ $S.m^{-1}$ at T > 750°C, **Fig.2b**). Iron increases garnet conductivity (Dai et al., 2012), due to the decrease of the average distance between $Fe^{2+}$ and $Fe^{3+}$ (Romano et al., 2006). It should be noted that similar values are obtained if the garnet composition is ~$Py_{73}$-$Alm_{14}$-$Grs_{13}$ (≈ 5.6 wt.% Fe; Dai et al., 2012) and $Py_{85}$-$Alm_{15}$ (≈ 6 wt.% Fe; Romano et al., 2006). In spite of the small size of the nucleated garnet crystals in the recovered samples, an approximate formula of ~$Py_{51}$-$Alm_{18}$-$Grs_{31}$ (≈ 6.9 wt.% Fe) is obtained using EDS measurements (**Methods, section M5**). Since garnet is surrounded by Fe-poor, Ca-rich clinopyroxenes, the almandine fraction (Alm) represents a compositional lower bound (i.e. the garnet iron content is ≥ 6.9 wt.%). Nonetheless, the conductivity of dry $Py_{20}$-$Alm_{76}$-$Grs_{4}$ (≈ 26.7 wt.% Fe) garnet does not exceed $10^{-3}$ $S.m^{-1}$ at 850°C (Dai et al, 2012), which is 100 times lower than the measured bulk conductivity value (**Fig.2b**). Therefore, the high conductivities of the pyroxenite samples cannot be solely attributed to the iron content of the garnet. Including minor water within garnets could, however, reproduce the observations. Although water is widely reported to be incorporated into pyroxenes, garnet can also accommodate a significant amount of water in its crystal lattice (Aines & Rossman, 1984; Maldener et al., 2003; **Suppl. text, section S1c**). The hydrogarnet defect is stable in pyrope at conditions relevant to the Tanzanian mantle (2.5 GPa, 1000°C; Ackermann et al., 1983). In the $MgO$-$Al_2O_3$-$SiO_2$-$H_2O$ system, water content in synthetic pyrope is about 500 ppm and consists of $(HO)_4^{4-}$ clusters (Ackermann et al., 1983). It reaches 620 ppm $H_2O$ in $Py_{26}$-$Alm_{20}$-$Grs_{54}$ from the Zagadochnaya kimberlite (Beran et al., 1993), up to 900 ppm $H_2O$ in pyrope beneath the Udachnaya Craton (Maldener et al., 2003) and 2500 ppm $H_2O$ beneath the Colorado Plateau (Aines & Rossman, 1984). Electrically, water in garnet significantly increases conductivity; for instance, hydrous pyrope (with 465 ppm $H_2O$) is 100 times more conductive at 750°C than its anhydrous counterpart (**Fig.2b**; Dai et al., 2012). Electrical models considering various connectivity modes (**Fig.2**, **Fig.S12, Fig.S13**) suggest that in the samples, bulk conductivity of pyroxenite (e.g. BB-246; **Fig.3b**) is best reproduced by the presence of a garnet-rich grain-boundary network containing connected graphite impurities. As a consequence, I suggest that the combination of a moderate amount of water, the presence of iron in garnet, and the preferential distribution of the garnet grains and graphitic impurities at the grain boundaries best explains the high electrical conductivity of the pyroxenite samples at 3 GPa.

Electrical anomalies in the deep Tanzanian lithosphere are reappraised in light of experimental results (**Fig.4**). Two endmember scenarios can be considered, depending on the amount and connectivity degree of garnets: the deep Tanzanian lithosphere is composed of either (1) fertile mantle rocks with garnet-rich networks or (2) depleted mantle rocks without garnet-rich networks. Garnet-rich networks can explain high-conductivity anomalies at depth > 60 km (P > 2 GPa). Shallow anomalies (< 70km) have been addressed by previous works and are discussed in the ***Suppl. text*** (***section S3***).

Garnet is ubiquitous in the deep cratonic lithosphere, but the conductive anomaly is located deeper than the top of the garnet stability field (**Fig.4a**). Furthermore, if the transition to the garnet stability field was solely the cause of the conductive anomaly, such a transition should be a ubiquitous feature within cratons globally, yet that is not the case. In Tanzania, the conductive anomaly corresponds to the deepest part of the cratonic lithosphere (the MT data are unable to resolve whether the anomaly crosses the LAB). The widespread presence of garnet at these depths is consistent with previous studies that argue for increased fertility with depth, associated with recent plume impingement (Lee & Rudnick, 1999). As a consequence, I propose that the garnet-rich networks originate from plume-induced metasomatism.

As illustrated in **Fig.4**, the experimental results on the pyroxenite and dunite at 3 GPa (**Fig.2b**) reproduce the field conductivity values reported by Selway (2015) for the deep lithosphere of the craton and the neighboring belt, respectively. I propose that the electrically conductive anomaly below the craton is caused by the presence of a garnet-rich network in a similar manner to the networks observed in the pyroxenite samples (**Fig.4e**). Scattered graphitic impurities, also introduced by plume metasomatism, may locally contribute to the electrical network. As previously noted (Selway, 2015), partial melting is not a plausible candidate to explain the field electrical observations, as both high shear-wave seismic velocities (O'Donnell et al., 2013) and a cool geotherm (Vauchez et al., 2005) have been reported for this craton. Instead, it has been suggested that the conductive anomaly could be due to a water content significantly higher in the cratonic mantle than in the highly deformed Mozambique Belt (Selway, 2015). This contradicts rheological studies highlighting that deformation is enhanced by the presence of water (Hirth & Kohlstedt, 2003), though differences in grain size could counter the weakening effect of large water content, allowing the cratonic root to remain stable (Selway, 2015).

The results of the present study also reconcile the apparent discrepancy between MT (high conductivity) and seismic observations (fast velocities) at 70 ±10 km depth (O'Donnell et al., 2013). The volume fraction of garnets required for the electrical model are not so substantial that a significant impact on bulk seismic velocity should be expected. It should be noted, however, that an enigmatic seismic discontinuity was evidenced beneath North America around these depths (Hales, 1969). Both the results and simulations

(**Fig.2**, **Fig.S12**, **Fig.S13**) recall that the connectivity of conductive minerals is the main parameter controlling bulk electrical conductivity, rather than their actual volume fraction. Hence, the garnet fraction needed to explain the electrical structure is sufficiently small that a commensurate change in seismic velocities is not expected (James et al., 2003). Although, the distribution of garnet-rich networks within the deep Tanzanian lithosphere is unknown, occurrences of garnet interconnected over long distances are observed in nature (***Suppl. text, section S1d***), for example within metamorphosed oceanic crust of relict subducted slabs (John et al., 2004). In our samples, the garnet network is likely assisted by minor amounts of graphite. Garnet networks, reported both in xenoliths (Henjes-Kunst & Altherr, 1992) and outcrops (Vrijmoed et al., 2013), can be caused by mantle refertilization.

Although conductive lithospheric mantle is not ubiquitous in cratons, there are examples of elevated lithospheric conductivity, including the rifted Yilgarn craton, Australia (Wang et al., 2014) and the Dharwar craton near the Deccan traps, India (Patro & Sarma, 2009). The North China craton is thought to have been disrupted by plume impingement (Wang et al., 2015) and has a signature that reflects thinner lithosphere after delamination (Ye et al., 2018). Metasomatism due to plume impingement has been suggested as a potential cause of the destruction of cratons (Foley, 2008; Wang et al., 2015), with the excess density within the root resulting from widespread garnet eventually leading to a gravitational instability and delamination (e.g. Foley, 2008). Elevated conductivities from garnets suggest extensive metasomatism by the African superplume, a model supported by numerical modelling for the East African rift system around Tanzania (e.g. Koptev et al., 2016). Thus, I suggest that the MT model of Tanzania (Selway et al., 2015) presents an image of lithosphere primed to break away, leading to a re-organization of the craton. Root delamination would eventually erase the conductive anomaly.

As for the deep cratonic lithosphere, I suggest in a recent study that the electrical structure of the oceanic lithospheric mantle of the Cocos Plate can also be explained by the presence of garnet pyroxenites with moderate hydration (Ferrand, submitted).

## Acknowledgments

I thank Emily Chin (UCSD-SiO GRD) for providing the starting materials, Rong Zhang and Danielle Shields for assistance in the laboratory, Rob Evans (WHOI) for key discussions on magnetotellurics, Christian Chopin for the constructive discussion on mineral identification and stability conditions, and Timm John for providing key references about garnet connectivity in natural rocks. I acknowledge SIO-PEPL for the facility and Kurt Leinenweber for providing cell assemblies; use of the COMPRES Cell Assembly Project was supported by COMPRES under NSF Cooperative Agreement EAR 1661511. I thank Jonathan Souders (UCSD-SiO IGPP) for the technical expertise on electrical circuits, Sabine Faulhaber (UCSD Nanoengineering

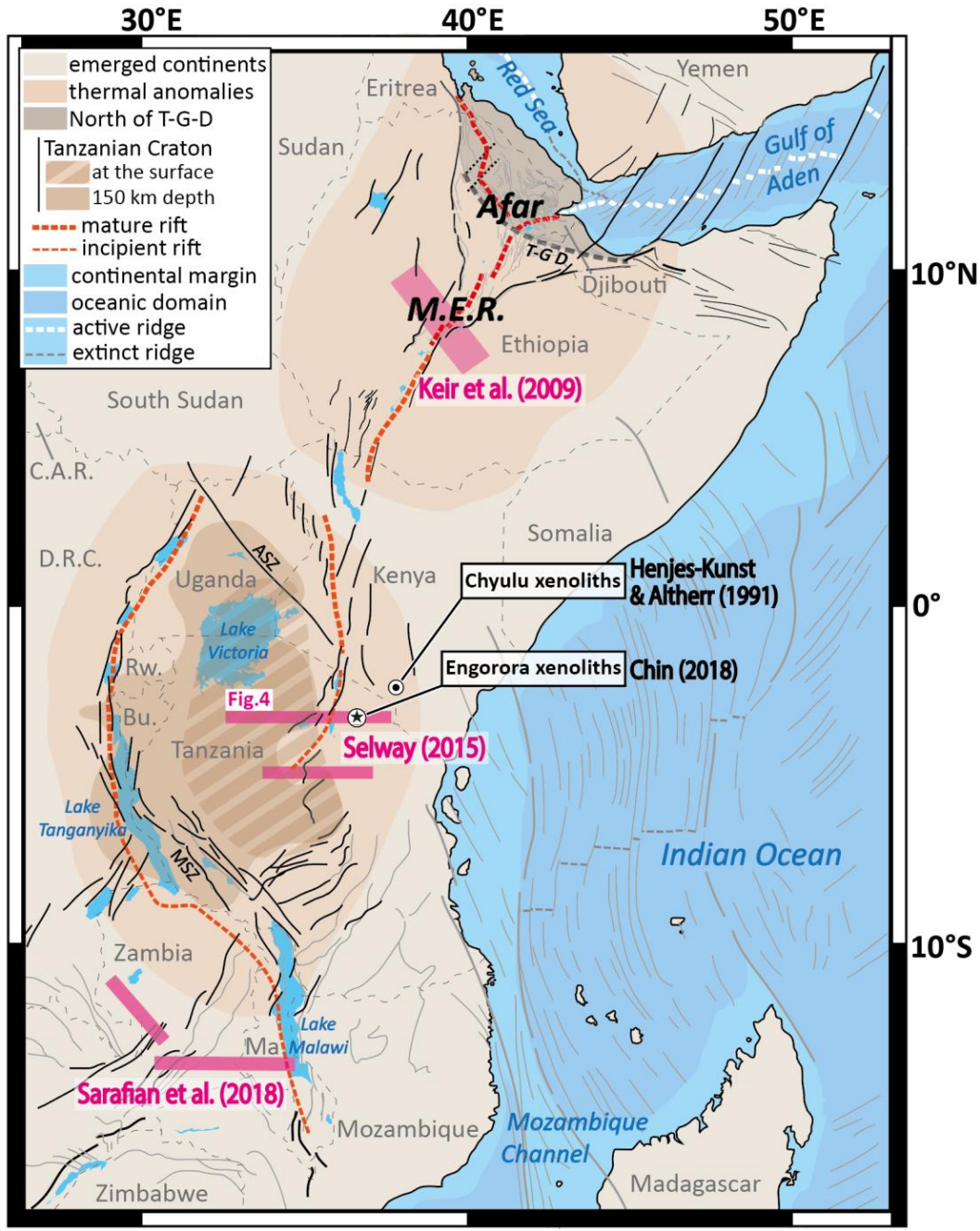

**Figure 1: Synthetic map of the East African Rift**, adapted from Chorowicz (2005), Sarafian et al. (2018) and Lemna et al. (2019), with oceanic fabric from Phethean et al. (2016). The limits of the Tanzanian Craton are from Koptev et al. (2016). The Blackstar indicates the origin of the Engorora xenoliths (Chin, 2018) used as starting material in this study; the black dot locates the origin of the Chyulu xenoliths (Henjes-Kunst & Altherr, 1992). Pink segments

locate the profiles of previous magnetotelluric studies across the Tanzanian (Selway, 2015) and Zambian (Sarafian et al., 2018) incipient rifts, and across the Main Ethiopan Rift (M.E.R.; Keir et al., 2009). Legend: AFZ: Aswa Fracture Zone; MFZ: Mughese Fracture Zone; T.G.D. = Tendaho-Goba'ad Discontinuity. Countries abbrv.: Bu. = Burundi; C.A.R. = Central African Republic; D.R.C. = Democratic Republic of the Congo; Ma. = Malawi; Rw. = Rwanda.

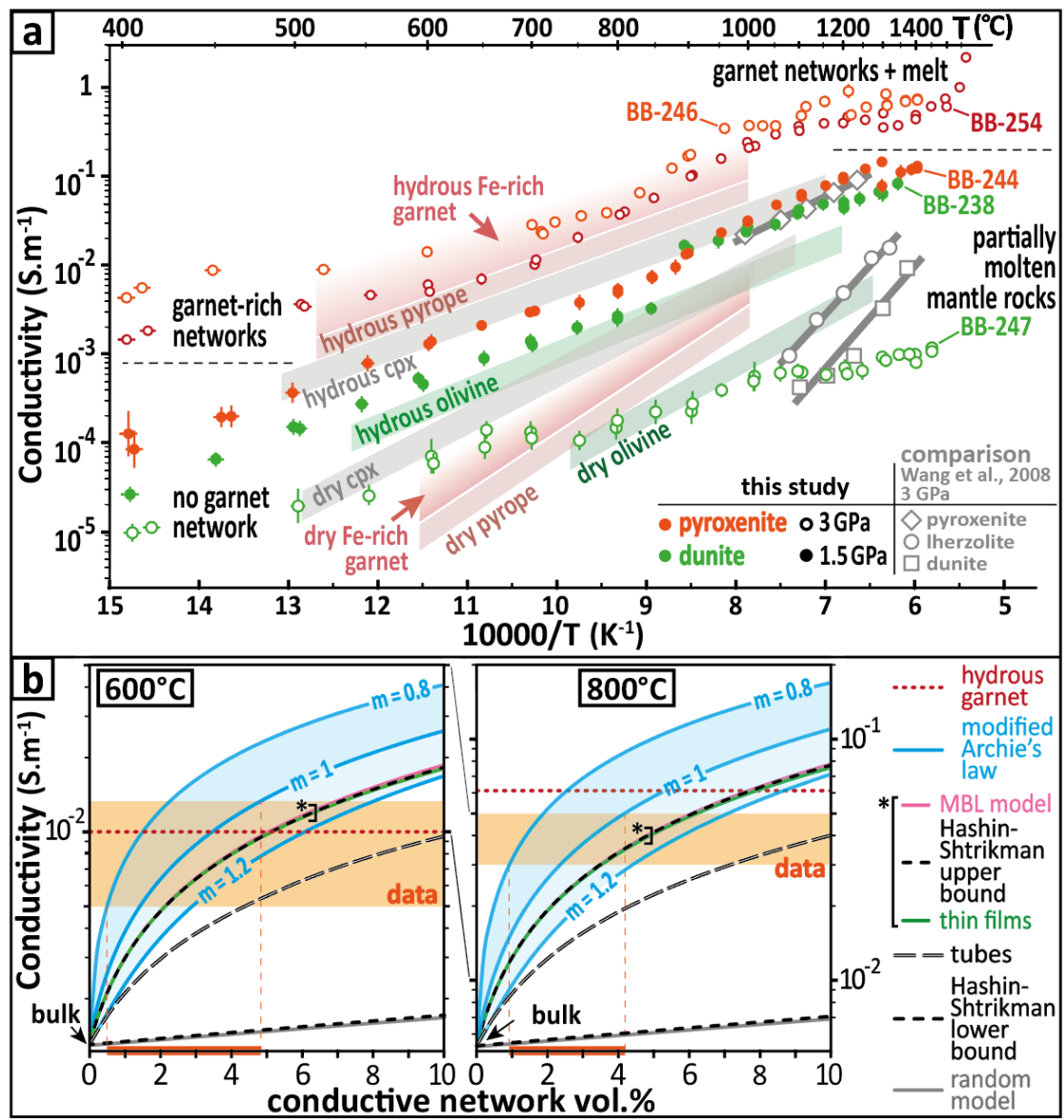

**Figure 2: Experimental results and comparison with models. (a)** Electrical conductivity versus reciprocal temperature for the Engorora xenoliths and a synthetic basaltic melt, at 1.5 GPa (solid circles) and 3 GPa (open circles). For comparison, the green, grey and red shades correspond to electrical conductivity values for dry ($X_{Fe}$ = 0.1, 6-19 GPa; Yoshino et al., 2012) and hydrous olivine (465 ppm $H_2O$; Gardés et al., 2014), dry and hydrous (375 ppm $H_2O$) clinopyroxene (Yang et al., 2011) and dry pyrope ($Py_{73}$-$Alm_{14}$-$Grs_{13}$, 3 GPa; Dai et al., 2012) and hydrous pyrope (465 ppm $H_2O$; 3 GPa; Dai et al., 2012). **(b)** conductivity simulations based on several models and compared to the experimental data at 600 and 800°C, i.e. before any melting. For details on the models, see **Methods** (**section M6**). For additional simulations, see **Fig.S12** and **Fig.S13**. The even higher conductivity observed in this study is attributed to hydrogarnet-rich Fe-rich pyrope. A summary of the experiments is provided in **Fig.S2**, **Table S2** and **Table S3**, including experiment BB-253 (ENG7, 3 GPa, ≤ 750°C), for which no electrical data was recorded. The parameters and different models are provided in **Table S6** to **Table S7**.

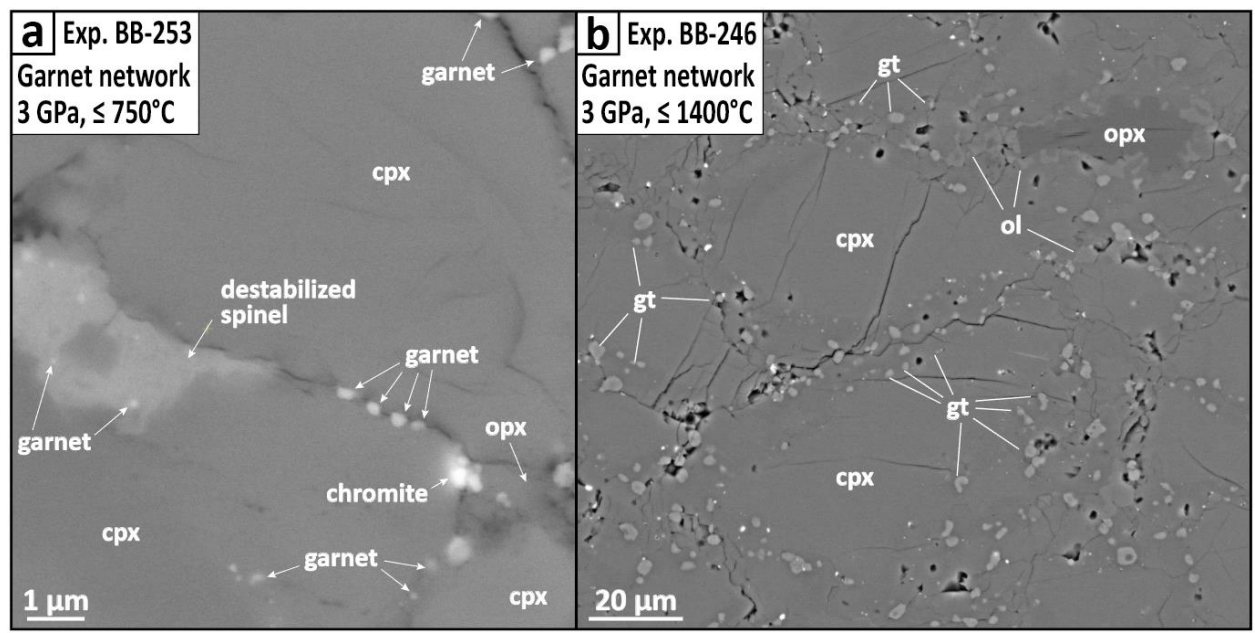

**Figure 3: High conductivity due to hydrous Fe-rich garnet networks at high-pressure. a:** Backscattered electron image (BSE) of an experimental garnet network (3 GPa, quench at 750°C), showing a destabilized spinel and submicrometric garnets crystallizing along grain boundaries; **b:** High-temperature garnet-rich channels (3 GPa, quench at 1400°C), showing larger garnet grains due to enhanced diffusion. For additional BSE images and EDS mapping, see **Fig.S6** to **Fig.S11**. During the spinel-garnet transition, Fe diffuses from destabilized spinel and oxides, while Al, Ca and OH are consumed from the surrounding minerals. Abbrev.: cpx = clinopyroxene; gt: garnet; ol = olivine; opx = orthopyroxene.

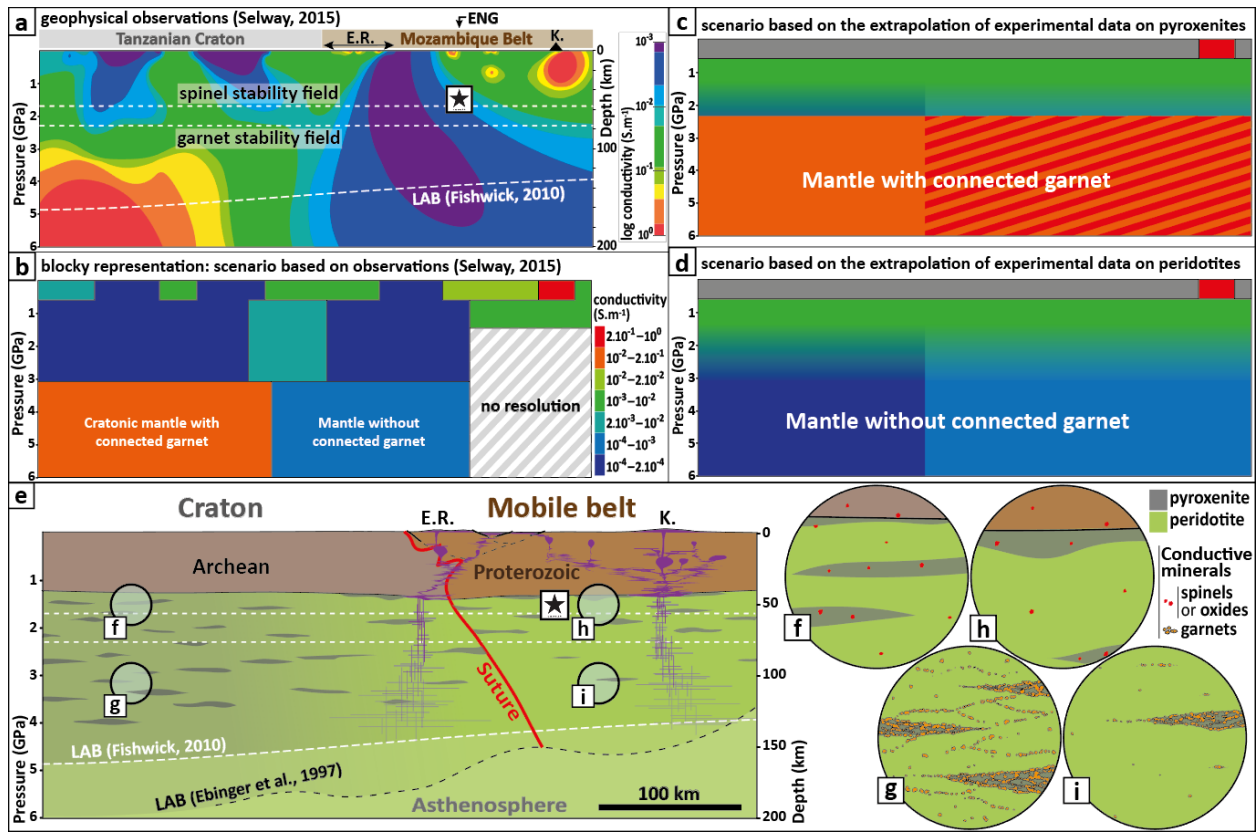

**Figure 4: Geophysical and geological implications.** (a) Results of the magnetotelluric survey (redrawn after Selway, 2015) located in **Fig.1**, with depth of the spinel-garnet transition (O'Hara et al., 1971) and LAB (Fishwisk, 2010); (b) Block representation of the best fit from connectivity data inversions (Selway, 2015); (c) block representation from the extrapolation of the experiments on pyroxenites; (d) block representation from the extrapolation of the experiments on pyroxenites; (e) Geological interpretation, with up-to-date topography (Wichura et al., 2011), Moho depth (Last et al., 1997; Julià et al., 2005), Lithosphere-Asthenosphere Boundary (LAB; Ebinger et al., 1997; Fishwick, 2010) and Eyasi suture zone (Ebinger et al., 1997) (vertical exaggeration x2). f) uppermost cratonic mantle at 1.5 GPa; (g) deep cratonic lithosphere containing a network of connected garnets; (h) uppermost mantle of the Mozambique Belt; (i) with limited garnet connectivity. The Blackstar indicates the approximate origin of the Engorora xenoliths (Chin, 2018), i.e. 0.9-1.7 GPa (**Fig.S4**). Legend: ENG = Engorora; E.R = Eyasi Rift; K = Kilimanjaro.

# References

1. Ackermann, L., Cemič, L. & Langer, K. 1983. Hydrogarnet substitution in pyrope: a possible location for "water" in the mantle. *Earth & Planetary Science Letters* **62**(2), 208-214.
2. Aines, R. D. & Rossman, G. R. 1984. Water content of mantle garnets. *Geology* **12**(12), 720-723.
3. Baptiste, V., Tommasi, A. & Demouchy, S. 2012. Deformation and hydration of the lithospheric mantle beneath the Kaapvaal craton, South Africa. *Lithos* **149**, 31-50.
4. Beran, A., Langer, K. & Andrut, M. 1993. Single crystal infrared spectra in the range of OH fundamentals of paragenetic garnet, omphacite and kyanite in an eklogitic mantle xenolith. *Mineralogy & Petrology* **48**(2-4), 257-268.
5. Bizimis, M. & Peslier, A. H. 2015. Water in Hawaiian garnet pyroxenites: Implications for water heterogeneity in the mantle. *Chemical Geology* **397**, 61-75.
6. Chin, E. J. 2018. Deep crustal cumulates reflect patterns of continental rift volcanism beneath Tanzania. *Contributions to Mineralogy & Petrology* **173**(10), 85.
7. Chin, E. J., Soustelle, V. & Liu, Y. 2020. An SPO-induced CPO in composite mantle xenoliths correlated with increasing melt-rock interaction. *Geochimica et Cosmochimica Acta* **278**, 199-218.
8. Chorowicz, J. 2005. The east African rift system. *Journal of African Earth Sciences* **43**(1-3), 379-410.
9. Dai, L., Li, H., Hu, H., Shan, S., Jiang, J. & Hui, K. 2012. The effect of chemical composition and oxygen fugacity on the electrical conductivity of dry and hydrous garnet at high temperatures and pressures. *Contributions to Mineralogy & Petrology* **163**(4), 689-700.
10. Doucet, L. S., Peslier, A. H., Ionov, D. A., Brandon, A. D., Golovin, A. V., Goncharov, A. G. & Ashchepkov, I. V. 2014. High water contents in the Siberian cratonic mantle linked to metasomatism: An FTIR study of Udachnaya peridotite xenoliths. *Geochimica et Cosmochimica Acta* **137**, 159-187.
11. Duba, A. G. & Shankland, T. J. 1982. Free carbon & electrical conductivity in the Earth's mantle. *Geophysical Research Letters* **9**(11), 1271-1274.
12. Ebinger, C., Djomani, Y. P., Mbede, E., Foster, A. & Dawson, J. B. 1997. Rifting archaean lithosphere: the Eyasi-Manyara-Natron rifts, East Africa. *Journal of the Geological Society* **154**(6), 947-960.
13. Evans, R. L., Jones, A. G., Garcia, X., Muller, M., Hamilton, M., Evans, S., ... & Hutchins, D. 2011. Electrical lithosphere beneath the Kaapvaal craton, southern Africa. *Journal of Geophysical Research: Solid Earth* **116**(B4).
14. Evans, R. L., Elsenbeck, J., Zhu, J., Abdelsalam, M. G., Sarafian, E., Mutamina, D., ... & Jones, A. G. 2019. Structure of the Lithosphere Beneath the Barotse Basin, Western Zambia, From Magnetotelluric Data. *Tectonics* **38**(2), 666-686.
15. Ferrand, T. P. submitted. Conductive channels in the deep oceanic lithosphere could consist of garnet pyroxenites at the fossilized Lithosphere-Asthenosphere Boundary. *Minerals*.
16. Ferrand, T. P. 2019. Seismicity and mineral destabilizations in the subducting mantle up to 6 GPa, 200 km depth. *Lithos* **334**, 205-230.

17. Fisher, J. C. 1951. Calculation of diffusion penetration curves for surface and grain boundary diffusion. *Journal of Applied Physics* **22**(1), 74-77.
18. Fishwick, S. 2010. Surface wave tomography: imaging of the lithosphere–asthenosphere boundary beneath central and southern Africa? *Lithos* **120**(1-2), 63-73.
19. Foley, S. F. 2008. Rejuvenation and erosion of the cratonic lithosphere. *Nature geoscience* **1**(8), 503-510.
20. Gardés, E., Gaillard, F. & Tarits, P. 2014. Toward a unified hydrous olivine electrical conductivity law. *Geochemistry, Geophysics, Geosystems* **15**(12), 4984-5000.
21. Hales, A. L. 1969. A seismic discontinuity in the lithosphere. *Earth & Planetary Science Letters* **7**(1), 44-46.
22. Henjes-Kunst, F. & Altherr, R. 1992. Metamorphic petrology of xenoliths from Kenya and northern Tanzania and implications for geotherms and lithospheric structures. *Journal of Petrology* **33**(5), 1125-1156.
23. Hirth, G. & Kohlstedt, D. 2003. Rheology of the upper mantle and the mantle wedge: A view from the experimentalists. Geophysical Monograph – American Geophysical Union **138**, 83-106.
24. Hirth, G., Evans, R. L. & Chave, A. D. 2000. Comparison of continental and oceanic mantle electrical conductivity: Is the Archean lithosphere dry? Geochemistry, Geophysics, Geosystems **1**(12).
25. James, D. E., Niu, F. & Rokosky, J. 2003. Crustal structure of the Kaapvaal craton and its significance for early crustal evolution. *Lithos* **71**(2-4), 413-429.
26. John, T., Scherer, E. E., Haase, K. & Schenk, V. 2004. Trace element fractionation during fluid-induced eclogitization in a subducting slab: trace element and Lu-Hf-Sm-Nd isotope systematics. *Earth & Planetary Science Letters* **227**(3-4), 441-456.
27. Jones, A. G., Fullea, J., Evans, R. L. & Muller, M. R. 2012. Water in cratonic lithosphere: Calibrating laboratory-determined models of electrical conductivity of mantle minerals using geophysical and petrological observations. *Geochemistry, Geophysics, Geosystems* **13**(6).
28. Julià, J., Ammon, C. J. & Nyblade, A. A. 2005. Evidence for mafic lower crust in Tanzania, East Africa, from joint inversion of receiver functions and Rayleigh wave dispersion velocities. *Geophysical Journal International* **162**(2), 555-569.
29. Keir, D., Bastow, I. D., Whaler, K. A., Daly, E., Cornwell, D. G. & Hautot, S. 2009. Lower crustal earthquakes near the Ethiopian rift induced by magmatic processes. *Geochemistry, Geophysics, Geosystems* **10**(6).
30. Koptev, A., Burov, E., Calais, E., Leroy, S., Gerya, T., Guillou-Frottier, L. & Cloetingh, S. 2016. Contrasted continental rifting via plume-craton interaction: Applications to Central East African Rift. *Geoscience Frontiers* **7**(2), 221-236.
31. Last, R. J., Nyblade, A. A., Langston, C. A. & Owens, T. J. 1997. Crustal structure of the East African Plateau from receiver functions and Rayleigh wave phase velocities. *Journal of Geophysical Research: Solid Earth* **102**(B11), 24469-24483.

32. Lee, C. T. & Rudnick, R. L. 1999. Compositionally stratified cratonic lithosphere: petrology and geochemistry of peridotite xenoliths from the Labait tuff cone, Tanzania. In *Proceedings of the 7th international Kimberlite conference* (pp. 503-521).

33. Lemna, O. S., Stephenson, R. & Cornwell, D. G. 2019. The role of pre-existing Precambrian structures in the development of Rukwa Rift Basin, southwest Tanzania. *Journal of African Earth Sciences* **150**, 607-625.

34. Maldener, J., Hösch, A., Langer, K., & Rauch, F. 2003. Hydrogen in some natural garnets studied by nuclear reaction analysis and vibrational spectroscopy. *Physics & Chemistry of Minerals* **30**(6), 337-344.

35. O'Donnell, J., Adams, A., Nyblade, A., Mulibo, G. & Tugume, F. 2013. The uppermost mantle shear wave velocity structure of eastern Africa from Rayleigh wave tomography: Constraints on rift evolution. *Geophysical Journal International* **194,** 961-978.

36. O'Hara, M. J., Richardson, S. W. & Wilson, G. 1971. Garnet-peridotite stability and occurrence in crust and mantle. *Contributions to Mineralogy & Petrology* **32**(1), 48-68.

37. Patro, P. K. & Sarma, S. V. S. 2009. Lithospheric electrical imaging of the Deccan trap covered region of western India. *Journal of Geophysical Research: Solid Earth* **114**(B1).

38. Pearson, D. G., Boyd, F. R., Haggerty, S. E., Pasteris, J. D., Field, S. W., Nixon, P. H. & Pokhilenko, N. P. 1994. The characterisation and origin of graphite in cratonic lithospheric mantle: a petrological carbon isotope and Raman spectroscopic study. *Contributions to Mineralogy & Petrology* **115**(4), 449-466.

39. Pernet-Fisher, J. F., Howarth, G. H., Pearson, D. G., Woodland, S., Barry, P. H., Pokhilenko, N. P., ... & Taylor, L. A. 2015. Plume impingement on the Siberian SCLM: Evidence from Re–Os isotope systematics. *Lithos* **218**, 141-154.

40. Peslier, A. H., Woodland, A. B., Bell, D. R. & Lazarov, M. 2010. Olivine water contents in the continental lithosphere and the longevity of cratons. *Nature* **467**(7311), 78-81.

41. Phethean, J. J., Kalnins, L. M., van Hunen, J., Biffi, P. G., Davies, R. J. & McCaffrey, K. J. 2016. Madagascar's escape from Africa: A high-resolution plate reconstruction for the Western Somali Basin and implications for supercontinent dispersal. *Geochemistry, Geophysics, Geosystems* **17**(12), 5036-5055.

42. Pommier, A., Leinenweber, K. & Tran, T. 2019. Mercury's thermal evolution controlled by an insulating liquid outermost core? *Earth & Planetary Science Letters* **517**, 125-134.

43. Rauch, M. & Keppler, H. 2002. Water solubility in orthopyroxene. *Contributions to Mineralogy & Petrology* **143**(5), 525-536.

44. Romano, C., Poe, B. T., Kreidie, N. & McCammon, C. A. 2006. Electrical conductivities of pyrope-almandine garnets up to 19 GPa and 1700 C. *American Mineralogist* **91**(8-9), 1371-1377.

45. Sarafian, E., Evans, R. L., Abdelsalam, M. G., Atekwana, E., Elsenbeck, J., Jones, A. G. & Chikambwe, E. 2018. Imaging Precambrian lithospheric structure in Zambia using electromagnetic methods. *Gondwana Research* **54**, 38-49.

46. Selway, K. 2015. Negligible effect of hydrogen content on plate strength in East Africa. *Nature Geoscience* **8**(7), 543-546.

47. Selway, K., Yi, J. & Karato, S-I. Water content of the Tanzanian lithosphere from magnetotelluric data: Implications for cratonic growth and stability. *Earth & Planetary Science Letters* **388**, 175-186 (2014).

48. Takahashi, E. 1990. Speculations on the Archean mantle: missing link between komatiite and depleted garnet peridotite. *Journal of Geophysical Research: Solid Earth* **95**(B10), 15941-15954.

49. Vauchez, A., Dineur, F. & Rudnick, R. 2005. Microstructure, texture and seismic anisotropy of the lithospheric mantle above a mantle plume: insights from the Labait volcano xenoliths (Tanzania). *Earth & Planetary Science Letters* **232**(3-4), 295-314.

50. Vrijmoed, J. C., Austrheim, H., John, T., Hin, R. C., Corfu, F. & Davies, G. R. 2013. Metasomatism in the ultrahigh-pressure Svartberget garnet-peridotite (Western Gneiss Region, Norway): implications for the transport of crust-derived fluids within the mantle. *Journal of Petrology* **54**(9), 1815-1848.

51. Wang, D., Mookherjee, M., Xu, Y. & Karato, S. I. 2006. The effect of water on the electrical conductivity of olivine. *Nature* **443**(7114), 977-980.

52. Wang, D., Li, H., Yi, L. & Shi, B. 2008. The electrical conductivity of upper-mantle rocks: water content in the upper mantle. *Physics & Chemistry of Minerals* **35**(3), 157-162.

53. Wang, D., Karato, S. I. & Jiang, Z. 2013. An experimental study of the influence of graphite on the electrical conductivity of olivine aggregates. *Geophysical Research Letters* **40**(10), 2028-2032.

54. Wang, L., Hitchman, A. P., Ogawa, Y., Siripunvaraporn, W., Ichiki, M. & Fuji-ta, K. 2014. A 3-D conductivity model of the Australian continent using observatory and magnetometer array data. *Geophysical Journal International*, **198**(2), 1143-1158.

55. Wang, H., van Hunen, J. & Pearson, D. G. 2015. The thinning of subcontinental lithosphere: The roles of plume impact and metasomatic weakening. *Geochemistry, Geophysics, Geosystems* **16**(4), 1156-1171.

56. Wichura, H., Bousquet, R., Oberhänsli, R., Strecker, M. R. & Trauth, M. H. 2011. The Mid-Miocene East African Plateau: a pre-rift topographic model inferred from the emplacement of the phonolitic Yatta lava flow, Kenya. *Geological Society, London, Special Publications* **357**(1), 285-300.

57. Yang, X., Keppler, H., McCammon, C., Ni, H., Xia, Q. & Fan, Q. 2011. Effect of water on the electrical conductivity of lower crustal clinopyroxene. *Journal of Geophysical Research: Solid Earth* **116**(B4).

58. Ye, G., Unsworth, M., Wei, W., Jin, S. & Liu, Z. 2019. The Lithospheric Structure of the Solonker Suture Zone and Adjacent Areas: Crustal Anisotropy Revealed by a High-Resolution Magnetotelluric Study. *Journal of Geophysical Research: Solid Earth* **124**(2), 1142-1163.

59. Yoshino, T., Shimojuku, A., Shan, S., Guo, X., Yamazaki, D., Ito, E., ... & Funakoshi, K. I. 2012. Effect of temperature, pressure and iron content on the electrical conductivity of olivine and its high-pressure polymorphs. *Journal of Geophysical Research: Solid Earth* **117**(B8).

60. Zhang, B. & Yoshino, T. 2017. Effect of graphite on the electrical conductivity of the lithospheric mantle. *Geochemistry, Geophysics, Geosystems* **18**(1), 23-40.

# Supplementary information – Garnet pyroxenites explain high electrical conductivity in the East African deep lithosphere

Thomas P. Ferrand[1,2,3]

1: Institute of Geophysics & Planetary Physics, Scripps Institution of Oceanography, UC San Diego, La Jolla, USA;
2: Institut des Sciences de la Terre d'Orléans, CNRS UMR 7327, Université d'Orléans, France;
3: Institüt für Geologische Wissenschaften, Freie Universität Berlin, Malteserstraße 74-100, Berlin 12249, Germany.*

*current location. Contact: *thomas.ferrand@fu-berlin.de*

## Methods:

### M1 - Starting materials

The xenoliths used for the experiments originate from Engorora, Northern Tanzania (Chin, 2018). The Engorora xenoliths consist of clinopyroxenites, wehrlites and dunites. In this study I used powders of a clinopyroxenite (ENG7) and a dunite (ENG8), which can be considered as two endmember mantle lithologies, hereafter referred to as fertile and depleted, respectively. The powders present similar particle size distributions, with mean values of about 12 and 9 µm, respectively (**Fig.S1**). Clinopyroxenites, the most abundant xenoliths, mostly consist of clinopyroxene (>85%), with minor olivine and orthopyroxene (≈ 10%) and accessory minerals such as spinel and chromite. These pyroxenites have a granular texture and a grain size between ≈ 1 mm and > 5 mm. Dunites exhibit coarser equigranular olivine (95%, ≈ 5 mm) with minor clinopyroxenes (4%). Compositions of ENG7, ENG8 and their host lava are provided in **Table S1**, along with compositions of their equivalent in Lashaine (Dawson et al., 1970) and of mantle rocks from a previous electrical conductivity study (Wang et al., 2008).

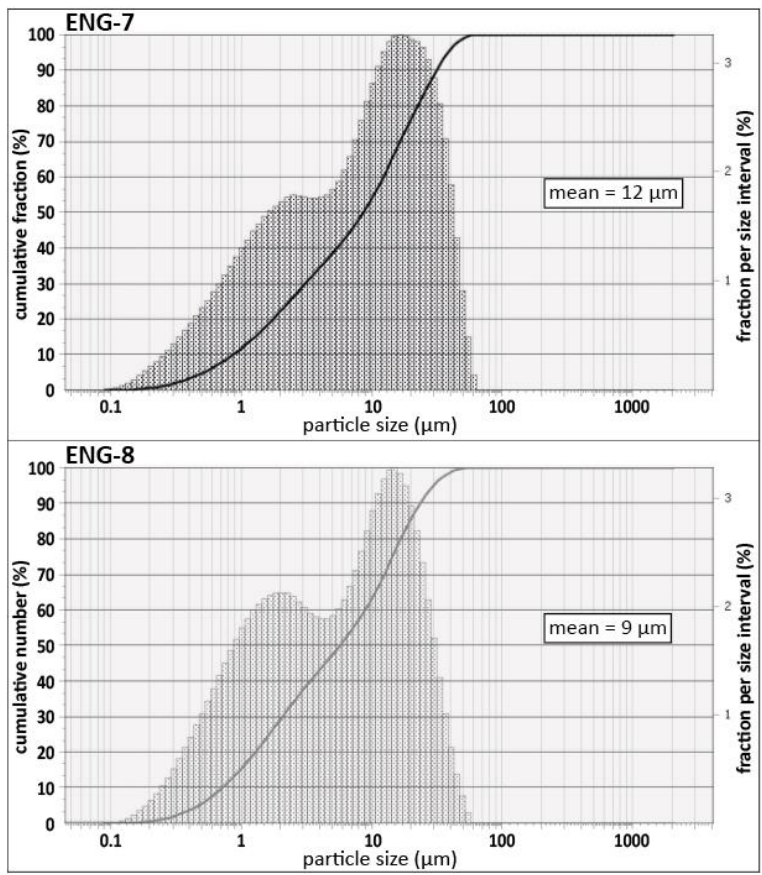

**Figure S1: Grain size distribution of the xenolith powders ENG7 and ENG8.** The xenoliths are thought to represent mafic cumulates or residua of rift-related magmatism (Chin, 2018). As Al-spinel is present in both ENG7 and ENG8, the equilibration pressures for these xenoliths are estimated to be between 0.9 and 1.7 GPa (Green & Hibberson, 1970). This pressure range corresponds to depths between 30 and 57 km, which is consistent with a location in the vicinity of the Moho.

The area is characterized by $CO_2$-rich magmatism (Dawson et al., 1970; Rudnick et al., 1993; Lee et al., 2000). This suggests that the studied xenoliths and the Tanzanian mantle contain minor amounts of graphite, i.e. the form of carbon expected for redox conditions typical for the uppermost mantle (***Suppl. text**, **section S2***).

Table S1: Compositions (XRF) of the Tanzanian and Chinese rocks on which electrical data exist.

| Location | Northern Tanzania | | | | | China | | |
|---|---|---|---|---|---|---|---|---|
| | Engorora | | | Lashaine | | Dengpa, Tibet | Wanquan, HeBei | |
| Sampling | Xenoliths | | | Xenoliths | | Outcrop | Xenoliths | |
| Nature | Dunite ENG-8 | Pyroxenite ENG-7 | Host lava | Lherzolite | Host lava | Dunite | Lherzolite | Pyroxenite |
| Major elements oxides (wt.%) | | | | | | | | |
| $SiO_2$ | 39.94 | 51.40 | 43.09 | 44.37 | 39.44 | 41.36 | 44.79 | 48.08 |
| $TiO_2$ | 0.16 | 0.57 | 2.47 | 0.08 | 2.37 | 0.01 | 0.08 | 0.62 |
| $Al_2O_3$ | 0.35 | 2.87 | 5.71 | 2.44 | 5.84 | 0.12 | 2.31 | 6.25 |
| $Cr_2O_3$ | 0.12 | 0.37 | 0.39 | 0.48 | 0.11 | n.d. | n.d. | n.d. |
| $FeO_T$ | 13.56 | 8.82 | 11.56 | 7.27 | 12.88 | 3.86 | 8.14 | 9.24 |
| $Fe_2O_3$ | n.d. | n.d. | n.d. | 0.85 | 9.21 | 0.34 | 1.89 | 3.04 |
| $FeO$ | n.d. | n.d. | n.d. | 6.42 | 4.59 | 3.52 | 6.25 | 6.20 |
| MnO | 0.19 | 0.18 | 0.20 | 0.09 | 0.17 | 0.04 | 0.13 | 0.15 |
| MgO | 43.87 | 17.66 | 18.71 | 42.14 | 17.67 | 52.34 | 40.72 | 17.27 |
| CaO | 0.91 | 17.09 | 12.35 | 1.45 | 12.24 | 0.95 | 2.26 | 15.43 |
| $Na_2O$ | 0 | 0.51 | 1.24 | 0.25 | 1.97 | 0.01 | 0.07 | 0.56 |
| $K_2O$ | 0.03 | 0.06 | 0.75 | 0.08 | 0.99 | 0.02 | 0.02 | 0.02 |
| $P_2O_5$ | 0.24 | 0.17 | 0.51 | 0.05 | 0.81 | 0.01 | 0.01 | 0.02 |
| $H_2O^+$ | n.d. | n.d. | 0.74 | 0.39 | 0.8 | 0.46 | 1.28 | 1.58 |
| $H_2O^-$ | n.d. | n.d. | n.d. | 0.18 | 2.26 | n.d. | n.d. | n.d. |
| $CO_2$ | n.d. | n.d. | 0.74 | 0.25 | 1.41 | 0.08 | 0.13 | 0.21 |
| Total | 99.38 | 99.69 | 96.98 | 99.52 | 99.88 | 98.72 | 98.53 | 97.64 |
| Mg# | 0.85 | 0.78 | 0.74 | 0.85 | 0.71 | 0.961 | 0.899 | 0.455 |
| Ref. | Chin, 2018 | | | Dawson et al., 1970 | | Wang et al., 2008 | | |

**M2 - Experimental protocol:**

All powders were stored in a desiccator to avoid water adsorption. Prior to experiments, the MgO parts of the assembly were fired to 1100°C during one hour and stored in a desiccator as well.

High-pressure, high-temperature experiments have been conducted using the 14/8 COMPRES electrical assembly (**Fig.S2**) designed for electrical conductivity measurements in the multi-anvil (Pommier & Leinenweber, 2018; Pommier et al., 2019). Each experiment was performed at a fixed pressure and the electrical conductivity was measured with increasing temperature, up to temperatures between 1400 and 1550°C depending on the experiment. **Table S2** summarizes the experimental conditions. An example of recovered quenched sample is presented in **Fig.S3**.

Each experiment on peridotites and pyroxenites reported in this study has been performed with the same protocol (see details in **Table S3**). First, pressure was increased from atmospheric pressure to the target pressure over several hours, then temperature was increased from room temperature to ≈ 400°C for a short dwell to achieve electrical equilibrium. The temperature was then lowered to < 300°C in order to start the experiment on an equilibrated system. Electrical measurements are then collected during heating until the sample is quenched at the highest T, except BB-238, for which a slow decrease of temperature was imposed after the peak temperature. Overall, the entire heating procedure until the target T of ≈ 400°C is reached and stable takes about 1h. During experiment BB-246, some instability in electrical measurements was noted upon heating (T < 800°C) after the dwell, and consequently the additional experiment BB-254 was performed. Such instability is interpreted as due to high-pressure transformations at grain boundaries in light of kinetics considerations. A discussion about the redox conditions and about the garnet network formation is provided in the *Suppl. text* (*section S2*).

To ensure reproducibility, experiment BB-246 has been reproduced with experiment BB-254 (**Table S2**). An additional experiment (BB-253) was performed on the pyroxenite powder up to 750°C only, i.e. quenched before any partial melting could occur, in order to check which microstructure was responsible for the drastic pressure-induced change in the conductivity values. No conductivity data were measured during this experiment. A summary of the experiments is provided in **Fig.2**. and **Table S2**. The experimental conditions are illustrated on **Fig.S4**.

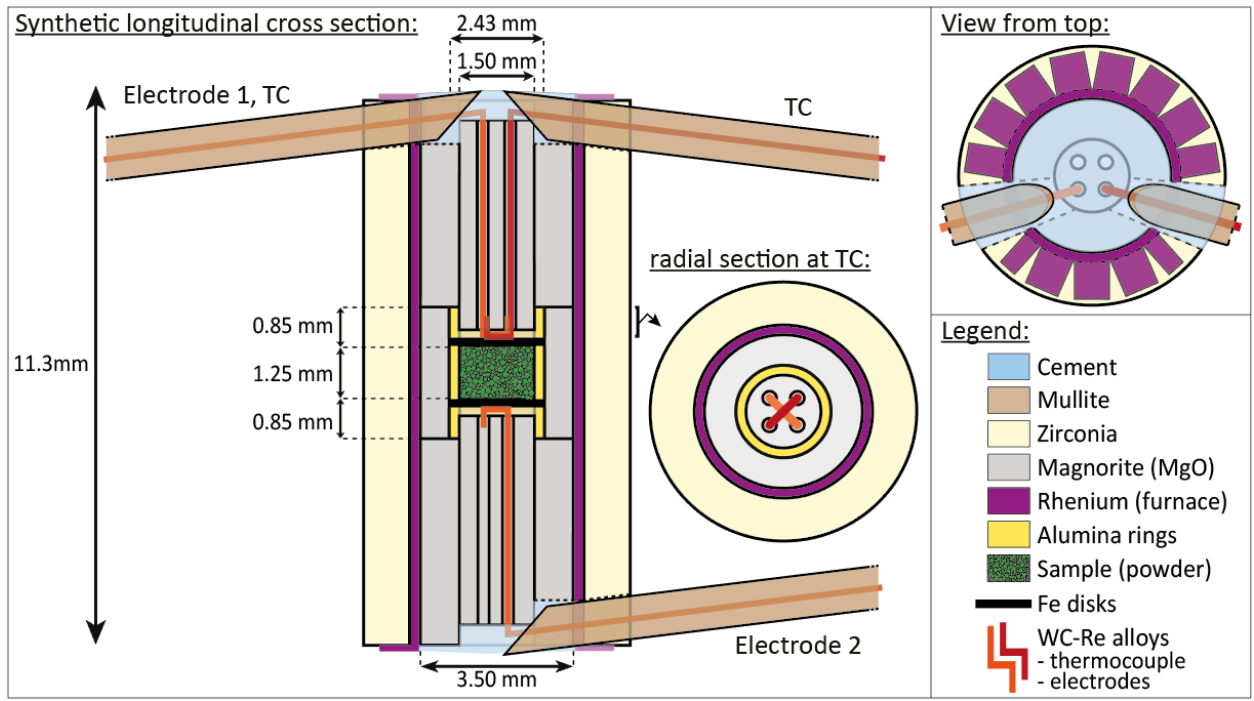

**Figure S2: Experimental assembly.** Synthetic sketch of the 14/8 COMPRES electrical assembly. Experiment BB-238 was performed using a 4-electrodes assembly (4 wires), which consist of a "symmetrical" assembly with 2 thermocouples.

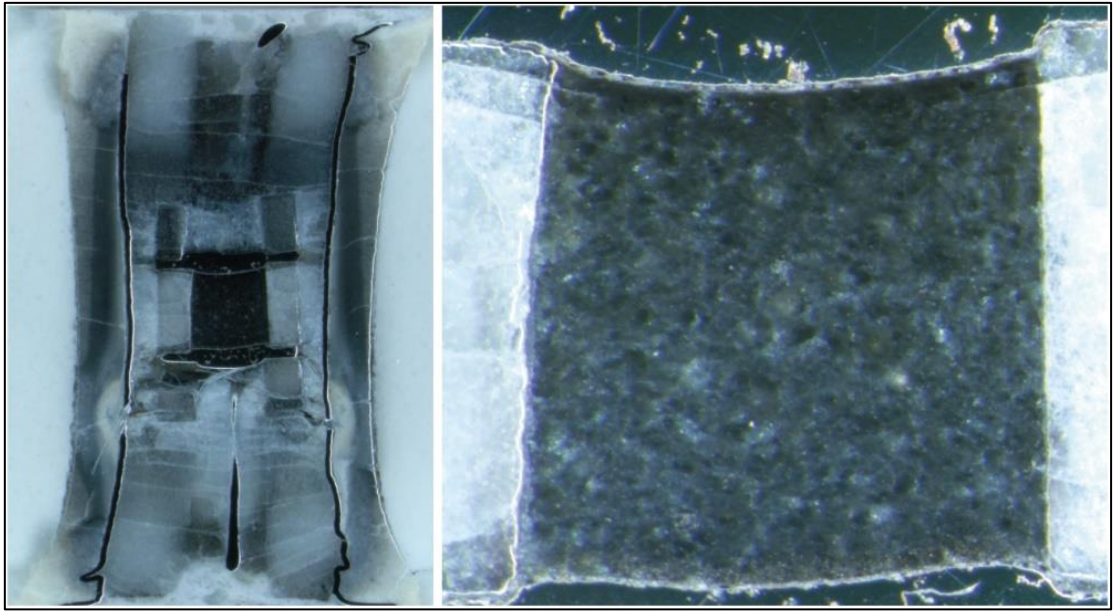

**Figure S3: Recovered assembly.** Run BB-246, clinopyroxenite ENG7.

**Table S2: Summary of the experiments.** For microstructural observations, see *section M5*.

| Run | Starting material | Pressure | Max. temperature | Microstructural observations |
|---|---|---|---|---|
| BB-244 | ENG7: pyroxenite xenolith powder | 1.5 GPa | 1403 °C | (Sub)micrometric spinel crystals localized in melt pockets. |
| BB-246 | | 3.0 GPa | 1400 °C | (sub)micrometric garnet networks. |
| BB-254 | | 3.0 GPa | 1575 °C | crystallized melt pockets containing Fe-rich dendrites (electrodes melting). |
| BB-253 | | 3.0 GPa | 750 °C | Submicrometric garnets at grain boundaries; no melting. |
| BB-238 | ENG8: dunite xenolith powder | 1.5 GPa | 1342 °C | melt wetting triple junctions. |
| BB-247 | | 3.0 GPa | 1449 °C | some garnet grains, not connected; few quenched silicate melt pockets. |

Table S3: Experimental details.

| Run | duration > 300°C | duration > 800°C | dwell temperature | dwell duration | stability check after dwell temperature range | stability check after dwell duration |
|---|---|---|---|---|---|---|
| BB-244 | 197 min | 107 min | 405 °C | 5 min | 405 °C ↘ 300 °C ↗ 403°C | 46 min |
| BB-246 | 241 min | 86 min | 402 °C | 5 min | 402 °C ↘ 179 °C ↗ 410°C | 48 min |
|  |  |  | 756 °C | 5 min | 756 °C ↘ 517 °C ↗ 786°C | 28 min |
| BB-254 | 156 min | 97 min | 402 °C | 5 min | 402 °C ↘ 202 °C ↗ 413°C | 20 min |
| BB-253 | 19 min | No electrical data due to broken wire → sample heated to 750°C for microstructures investigation | | | | |
| BB-238 | 165 min | 86 min | 400°C | 16 min | 400 °C ↘ 300 °C ↗ 400°C | 32 min |
| BB-247 | 178 min | 81 min | 410-420 °C | 11 min | 420 °C ↘ 201 °C ↗ 403°C | 43 min |

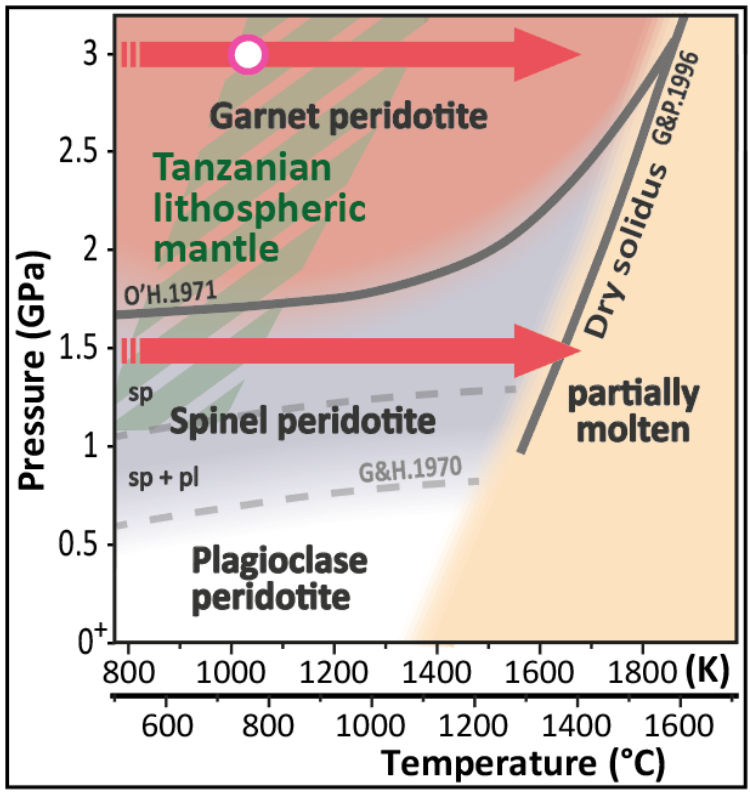

**Figure S4: Experimental conditions.** Pressure-temperature diagram summarizing experimental conditions (red arrows) and highlighting the spinel-garnet transition in mantle rocks and the Tanzanian lithospheric mantle. The white dot indicates the maximum temperature (750°C) during experiment BB-253 (ENG7, 3 GPa; no elec. data). Legend: pl = plagioclase; sp = spinel. Ref: G&P.1996 = Gudfinnsson & Presnall (1996); O'H.1971 = O'Hara et al. (1971); G&H.1970 = Green & Hibberson (1970).

## M3 - Conductivity measurements and uncertainties

The electrical cell assembly is shown in **Fig.S2**. In this assembly, the starting powder was packed into an alumina sleeve, and both the powder and two Fe electrode disks are held together in a three-alumina ring system. The rings are surrounded by high-purity MgO sleeves, a rhenium furnace, and a zirconia insulating sleeve. The whole column is inserted in an MgO-spinel pressure medium. W-Re wires are used as both electrodes and thermocouples (Pommier et al., 2019). A current with a controlled voltage (DC potential of 1 V and AC amplitude between 500 and 1000 mV) is applied during the measurement. The complex impedance was measured using an impedance analyzer (1260 Solartron Impedance/Gain-Phase Analyzer) and was directly collected over a frequency range between $10^6$ Hz and $10^{-1}$ Hz. As illustrated in **Fig.S5**, the electrical cell provides high-quality spectra over wide frequency ranges, as expected for such material in these conditions (Pommier & Leinenweber, 2018).

The resistance of the sample corresponds to the real part of the complex impedance spectrum (**Fig.S5**). Because I used the 2-electrodes (3 wires) assembly, 7.5 Ω were subtracted from the measured resistance values, corresponding to the contribution of the electrode wires and BNC cables, to the overall resistance of the electrical circuit. This correction does not apply for experiment BB-238, for which a 4-electrode assembly (4 wires) was used (Pommier & Leinenweber, 2018).

The conductivity $\sigma$ is calculated from $R$ using the following formula:

$$\sigma = \frac{l}{(R.A)}$$

where $l$ is the height of the sample (≈ 1.2 mm) and $A$ is the sample area (≈ 1.9 .$10^{-6}$ m²).

Errors on the electrical conductivity data $\Delta\sigma$ are the total of the errors on l, A and $R$, and calculated as follows:

$$|\Delta\sigma| = \frac{\Delta l}{|R.A|} + \left|\frac{l}{A.R^2}\right|\Delta R + \left|\frac{l}{R.A^2}\right|\Delta A$$

In **Fig.2a**, these errors are, for most, smaller than the size of the datapoints.

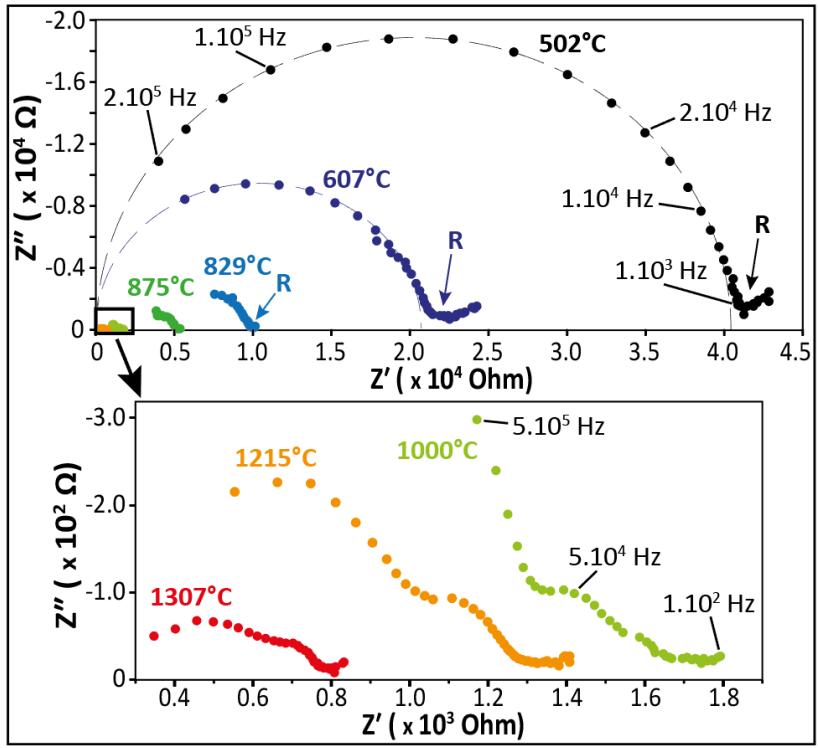

Figure S5: Examples of impedance spectra collected using the electrical cell (2-electrode measurements) during experiment BB-246 (pyroxenite). Z' and Z'' are the real and imaginary part of the complex impedance, respectively. Intersection of the sample's response with the real axis provides the value of the bulk electrical resistance. The dashed ellipses highlight that the shape of the measured spectra is close to the theoretical shape.

### M4 - Temperature measurements and uncertainties

The temperature was monitored with a $W_{95}Re_5$–$W_{74}Re_{26}$ (C-type) thermocouple inserted within a 4-bore MgO sleeve with the junction in contact with the top of one of the two iron foils.

During experiment BB-238, the two thermocouples indicated a temperature difference lower than 5°C. As a consequence, and consistently with previous studies using the same setup, I considered a maximum temperature uncertainty of ±5°C, i.e. smaller than the size of the datapoints on **Fig.2a**.

### M5 - Microstructural observations using Scanning Electron Microscopy (SEM)

Retrieved assemblies were cut along the furnace and polished using sandpaper and diamond paste down to 1 µm. Microstructures (**Fig.S6**, **S7**, **S8** and **S9**) were imaged using a scanning electron microscope (SEM) at UCSD, Nanoengineering Facility. Mineral composition (**Table S4**) and element mapping (**Fig.S10** and **S11**) were acquired using energy dispersive X-ray spectroscopy (EDS).

Counting pixels using *Adobe Photoshop* on **Fig.3b** (3GPa; ≤1400°C), the modal abundance of garnet is estimated to be a few percent, e.g. ~5 vol.% for run BB-246. A similar analysis on **Fig.S7a** (large garnet cluster around metal-rich inclusions) gives garnet fraction estimates of 7-8 vol.%, but the occurrence of local clusters cannot increase the overall connectivity of the conductive garnet-rich network.

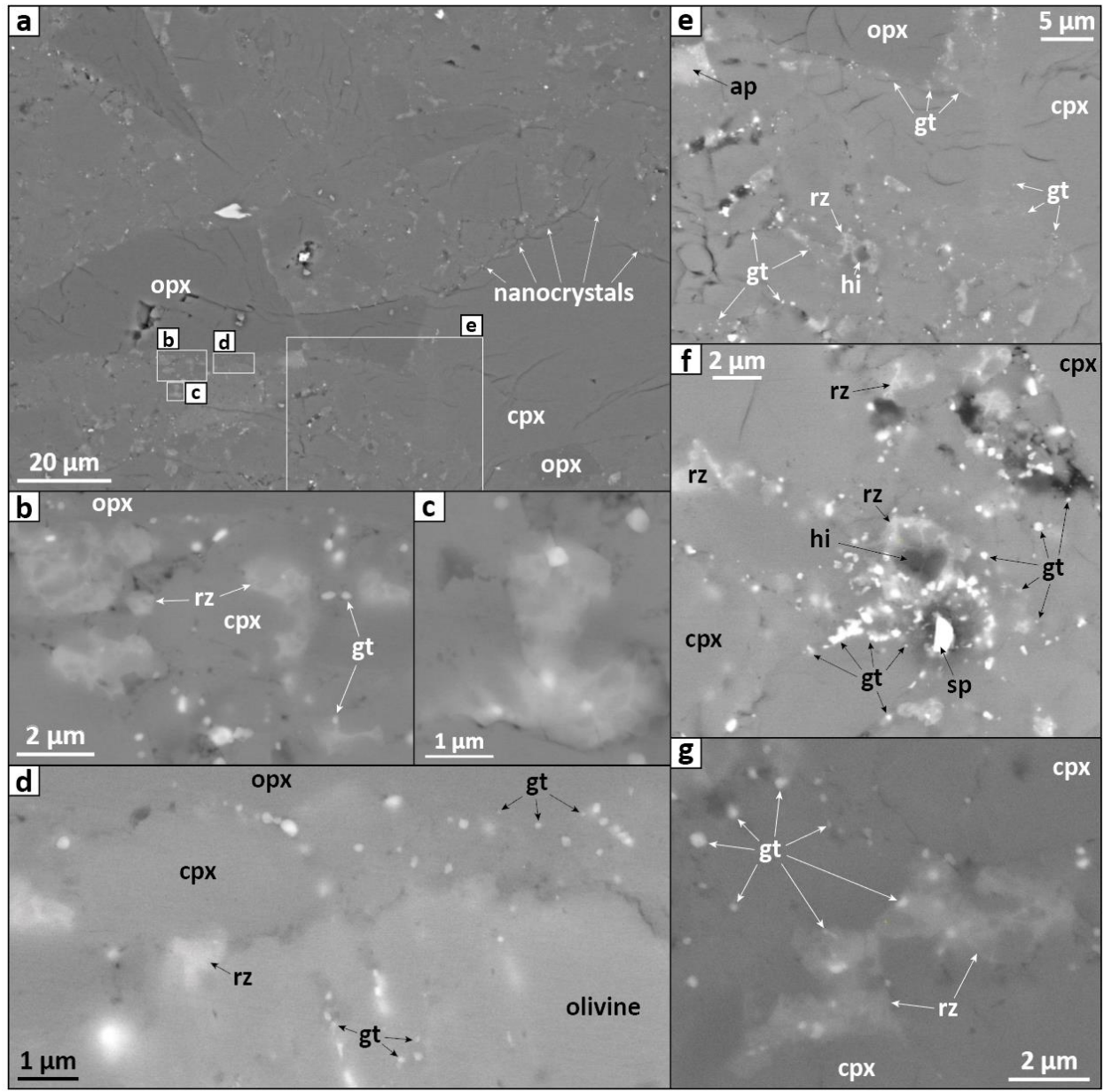

**Figure S6: Additional BSE images on sample BB-253 (pyroxenite, 3 GPa, 750°C).** Supplement of **Fig.3a**. Abbrev.: cpx = clinopyroxene; gt: garnet; hi = hibonite; opx = orthopyroxene; rz = reaction zone; sp = spinel.

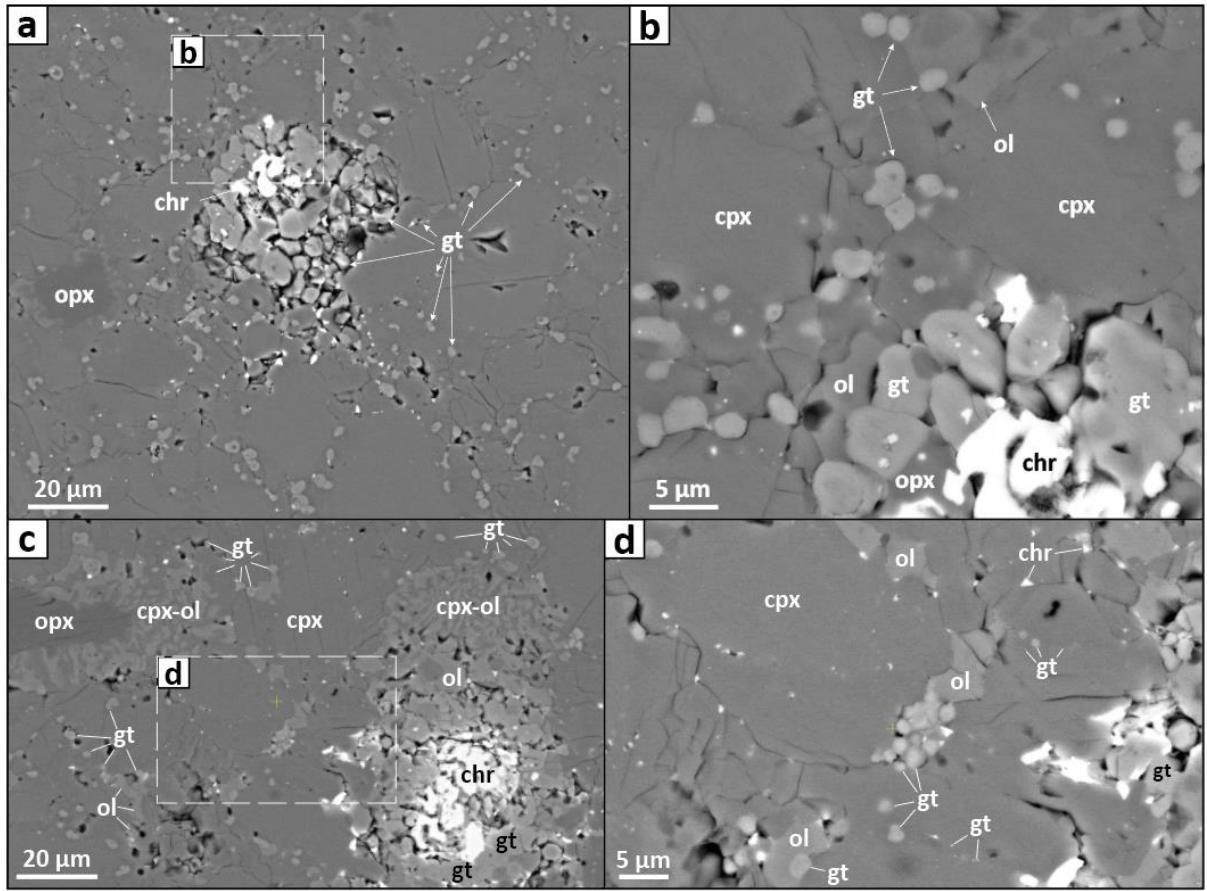

**Figure S7: Additional BSE images on sample BB-246 (pyroxenite, 3 GPa, 1400°C).** Supplement of **Fig.3b**. Networks of micrometric garnet crystals develop along grain boundaries throughout the sample. Larger garnet grains grow at the expense of Al-rich spinel, while chromite crystals remain stable. In addition, symplectites develop between opx, cpx and accessory minerals, which highlight metamorphic reactions. Abbrev.: chr: chromite; cpx = clinopyroxene; ol = olivine; opx = orthopyroxene; gt = garnet.

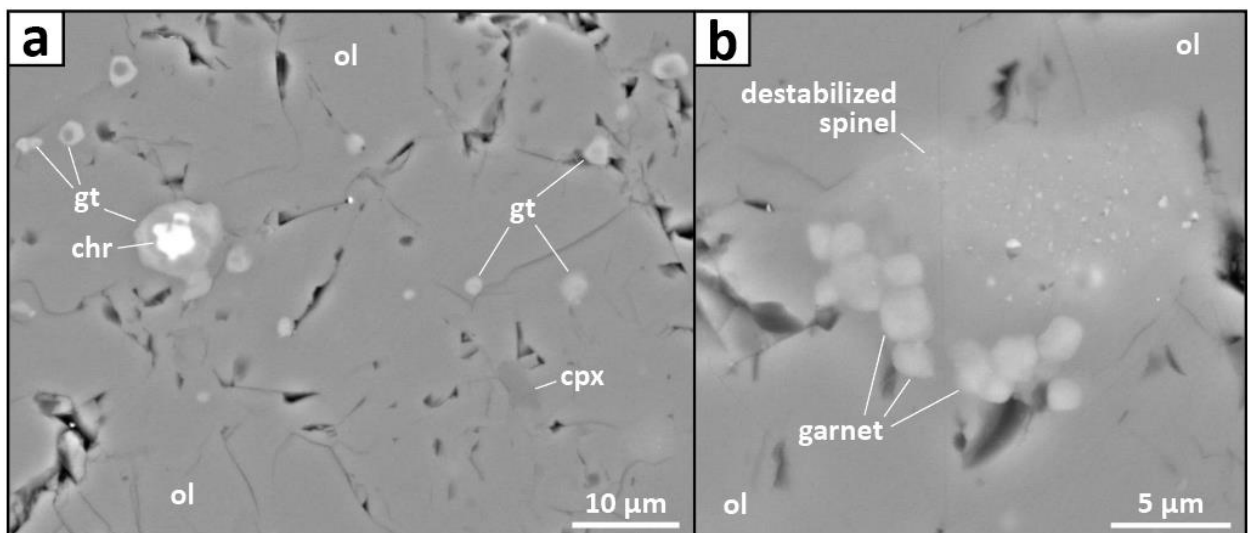

**Figure S8: Additional BSE images on sample BB-247 (dunite, 3 GPa, 1449°C).** Micrometric garnet grains grow only locally, in the vicinity of minor metal oxides (**a**) and destabilized spinel (**b**). Abbrev.: cpx = clinopyroxene; ol = olivine; chr: chromite; gt = garnet.

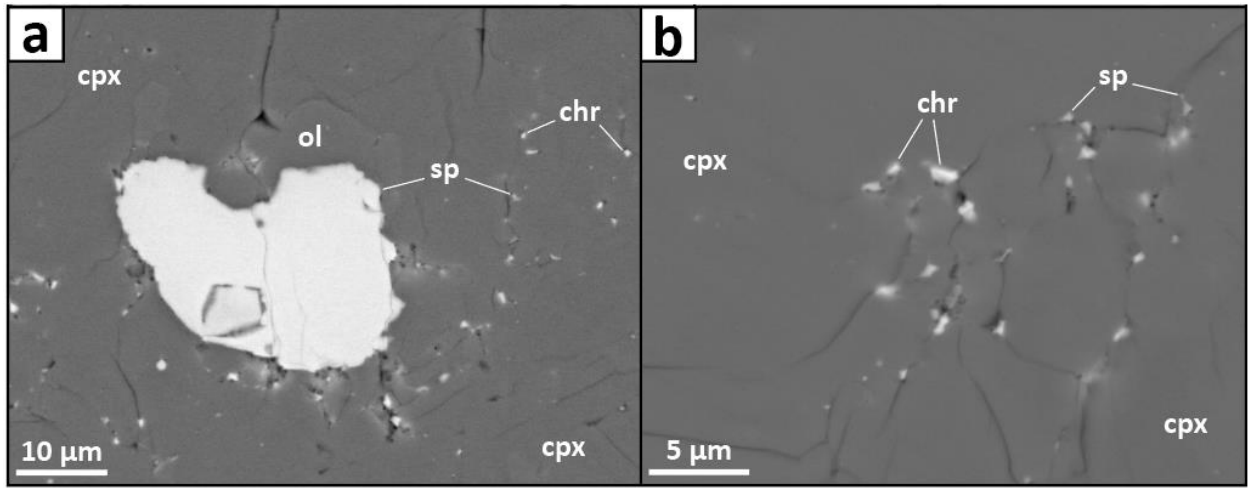

**Figure S9: Additional BSE images on sample BB-244 (pyroxenite, 1.5 GPa, 1403°C).** No garnet could develop at 1.5, i.e. garnet stability field (**Fig.S4**). Both chromite and Al-rich spinel, originating from the xenolith powder, remain stable. Abbrev.: ol = olivine; cpx = clinopyroxene; chr: chromite; sp = spinel.

**Table S4: EDS data from sample BB-253.** See analysis located on **Fig.S11**. *ND*: not detected.

| Element (wt.%) | Spectrum 1 | Spectrum 2 | Spectrum 3 | Spectrum 4 | Spectrum 5 |
|---|---|---|---|---|---|
| B | 0.86 | n.d. | n.d. | n.d. | n.d. |
| O | 36.42 | 38.97 | 39.11 | 38.98 | 40.45 |
| Na | n.d. | 0.47 | n.d. | 0.45 | n.d. |
| Mg | 25.12 | 10.35 | 8.32 | 10.26 | 18.05 |
| Al | 0.01 | 1.62 | 6.21 | 1.65 | 0.96 |
| Si | 18.98 | 25.75 | 21.99 | 25.72 | 27.13 |
| P | 0.05 | 0.05 | 0.53 | 0.05 | 0.07 |
| Ca | 0.21 | 14.51 | 11.36 | 14.75 | 1.52 |
| Ti | 0.02 | 0.33 | 0.48 | 0.37 | 0.17 |
| V | n.d. | 0.13 | n.d. | n.d. | n.d. |
| Cr | 0.04 | 0.54 | 0.49 | 0.6 | 0.22 |
| Mn | 0.21 | 0.09 | 0.21 | 0.07 | 0.2 |
| Fe | 15.75 | 6.15 | 10.47 | 6.12 | 9.65 |
| Co | 0.29 | n.d. | n.d. | n.d. | n.d. |
| Ni | 0.19 | 0.03 | 0.01 | n.d. | 0.01 |
| Lu | 0.48 | n.d. | n.d. | n.d. | n.d. |
| Ta | 0.47 | 0.37 | 0.35 | 0.3 | 0.5 |
| Tl | 0.31 | 0.31 | 0.31 | 0.33 | 0.44 |
| Th | 0.59 | 0.33 | 0.17 | 0.34 | 0.63 |

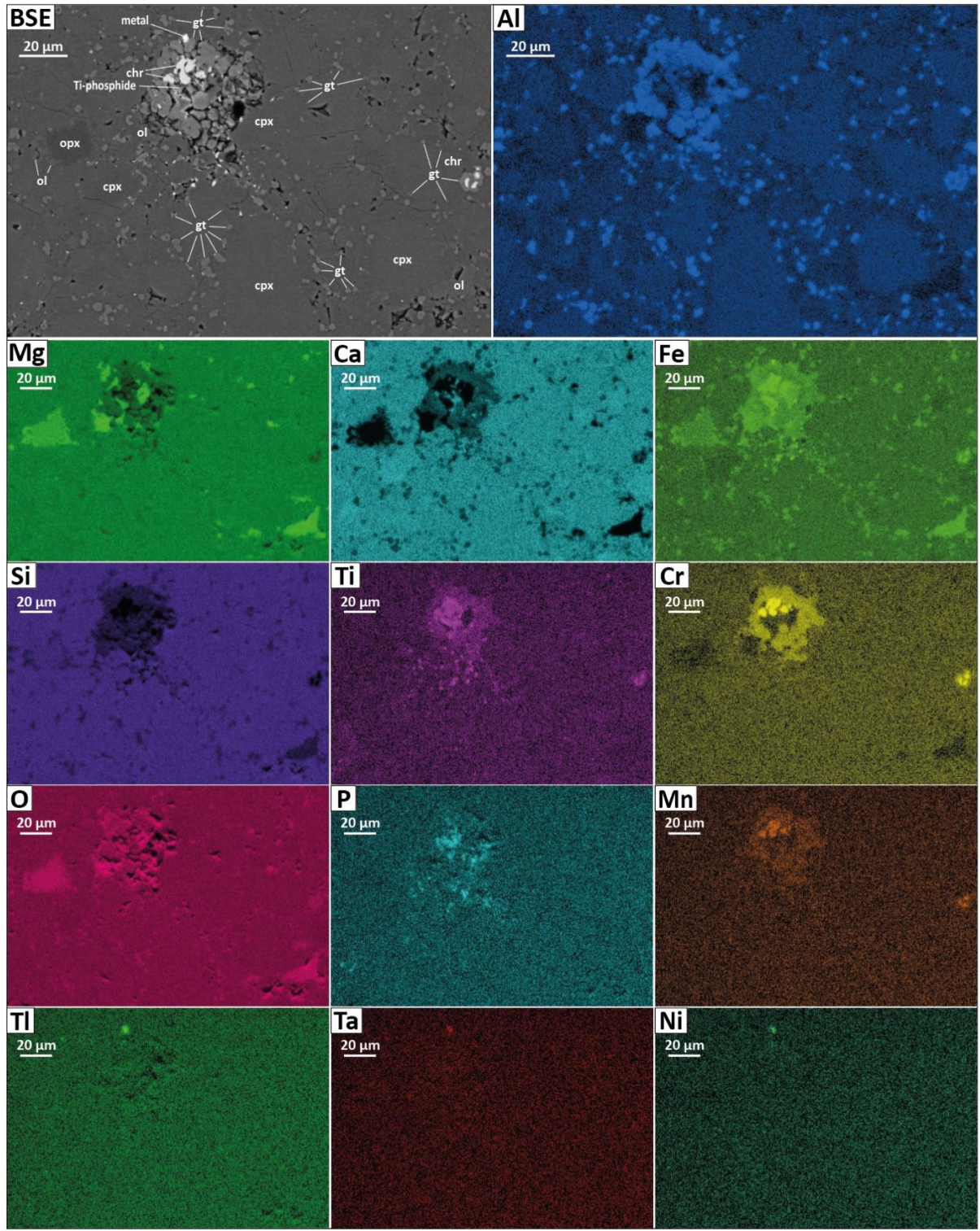

**Figure S10: Electron Dispersive X-ray spectroscopy (EDS) mapping on sample BB-246 (pyroxenite, 3 GPa, 1400°C).** Compositional and fugacity aspects are discussed in the *Suppl. text* (*section S2*). NB: these element maps are constructed from the relative peak intensities, which may significantly change with crystal structure (e.g. opx vs cpx, which contain the same amount of Si and O, contrary to what appears for the O map).

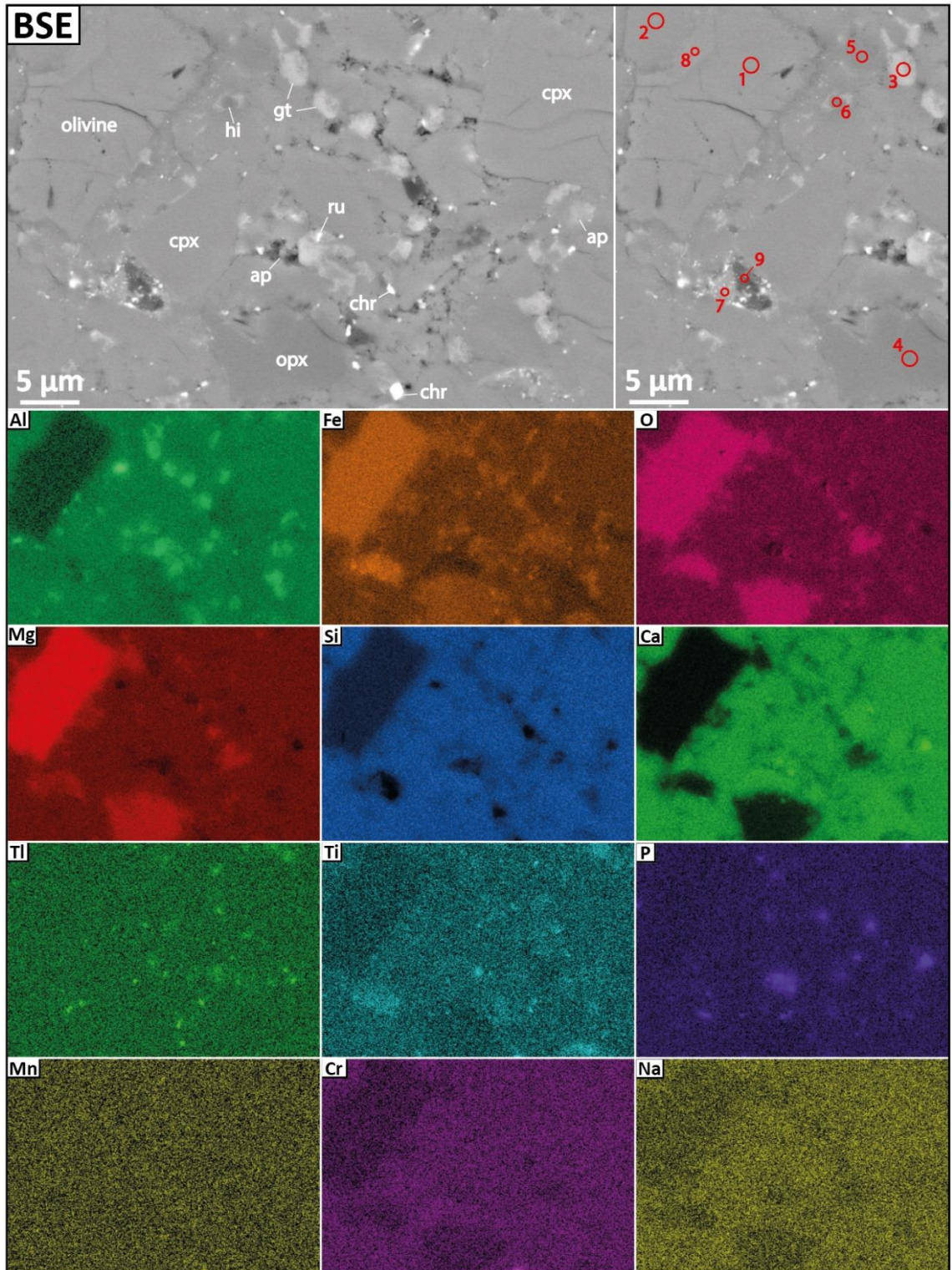

**Figure S11: Electron Dispersive X-ray spectroscopy (EDS) mapping on sample BB-253 (pyroxenite, 3 GPa, 750°C).** Abbrev.: ap = apatite; chr: chromite; cpx = clinopyroxene; gt = garnet; hi = hibonite; opx = orthopyroxene; ru =rutile; sp = spinel. For examples of EDS data, see **Table S4**.

## M6 - Network connectivity and electrical conductivity

The interconnection of conductive phases and the topology of the conductive network control the bulk electrical conductivity of a multiphase system (e.g. Glover et al., 2000; Wang et al., 2013; Miller et al., 2015). In the system considered in this study, we have to consider two networks which themselves combine to form the bulk conductivity. The first of these networks consists of the various minerals that compose the host rock. The second is the material within the grain boundaries that contains a complex distribution of phases, which I refer to as the intergranular system.

The starting material that forms the host is a relatively homogeneous distribution of submicrometric to plurimicrometric grains of olivine, pyroxenes and accessory minerals. I used a random-distribution model to compute the bulk conductivity of the host rock based on the documented conductivity data of each phase. Considering a harmonic mean with arbitrary shaped and oriented volumes of conductivity $\sigma_i$ and fraction $X_i$, a random model can be used to calculate the bulk conductivity $\sigma^*$ as follows:

$$\sigma^* = \prod_{i=1}^{N} \sigma_i^{X_i}$$

The conductivity within the grain-boundary network is less straightforward, as it consists of garnets, graphite and other phases distributed in a complex manner. However, when considering the effects of minor graphite impurities (**Suppl. text, section S2**), and given that there is no reason for graphite to be anisotropically distributed in the powder, the random model can also be used to estimate the conductivity of the graphite-bearing garnet-rich network. An alternative estimate would be to consider graphite impurities as electrical shorts between the garnet grains. The random model provides a more conservative estimate and is chosen for bulk calculations.

It should be noted that, considering graphitic impurities at grain boundaries, the graphite content of the grain-boundary conductive network is necessarily significantly larger than the bulk graphite fraction. While the bulk graphite content is estimated to be $\sim$ 0.1 wt.% (**Suppl. text, section S2**), I expect, considering the intergranular system equal to 3-10% of the total volume, that the average graphite content of the grain boundaries is < 10-30 vol.%. In **Fig.2** and **Fig.S12**, a garnet/graphite ratio of 80:20 is assumed, and **Fig.S13** also presents simulations for ratios 70:30 and 90:10. Electrical conductivity values for given fractions of garnet and graphite are provided in **Table S5**.

Several models have been developed to predict the bulk conductivity of rocks as a function of a conductive interstitial phase, typically partial melt or saline fluid (e.g. Glover et al., 2000; ten Grotenhuis et al., 2005, and references therein). Connectivity problems can be solved using the percolation theory (Gueguen & Dienes, 1989; Stauffer & Aharony, 1992). The connectivity threshold depends on the geometry of the conductive network (e.g. Miller et al., 2015). In many ways, the geometry addressed here is analogous to the melt-grain networks analyzed by Zhu et al. (2011) and Miller et al. (2015), except that instead of a homogenous melt phase (with a single conductivity value), we have a multi-phase network between grains that includes graphite and garnet as discussed above.

Archie's law (Archie, 1942) is a commonly used relationship to compute bulk conductivity that, in its most general form, relies only on knowledge of the conductive material between the grains and the geometry of that network. In Archie's law, the value of an exponent $m$ implicitly contains the effective connectivity of the conductive phase. A modified form of the expression accounts for the conductivity of the grains themselves (Glover et al. 2000) **(Table S6)**. Miller et al., (2015) found that a modified Archie's law is relevant to interpret bulk conductivity values as a function of melt fraction. Using microtomography in partially molten rocks (Zhu et al., 2011; Miller et al., 2015), it was observed that the connectivity threshold strongly differs between 2-D and 3-D systems (e.g. Clerc et al., 1979), which implies that the expected connectivity level in the samples is necessarily higher than what can be observed with SEM imaging (**Fig.3**, **Fig.S6**, **Fig.S7**).

I have used a variety of multi-phase conductivity models shown in **Table S6** in order to cross-check the bulk conductivities measured in the laboratory. Volume fraction of conductive phases was varied and preferred models correspond to the ones that reproduce the observations. As illustrated in **Fig.S12**, considering a moderate garnet hydration and the presence of a minor amount of graphite, the conductive grain-boundary network is expected to represent only 1 – 3 vol.%. Regarding the geometry of the garnet-rich network, it appears that either the Hashin-Shtrikman upper bound, the wetted thin films or the modified brick-layer models provide simulations that fit best the data (**Fig.S12**). Considering the uncertainty regarding both the graphite fraction and the exact water content of garnet, additional simulations show that the conductivity values obtained during the experiments at 3 GPa can be explained by the nucleation of nanometric garnet crystals at grain boundaries.

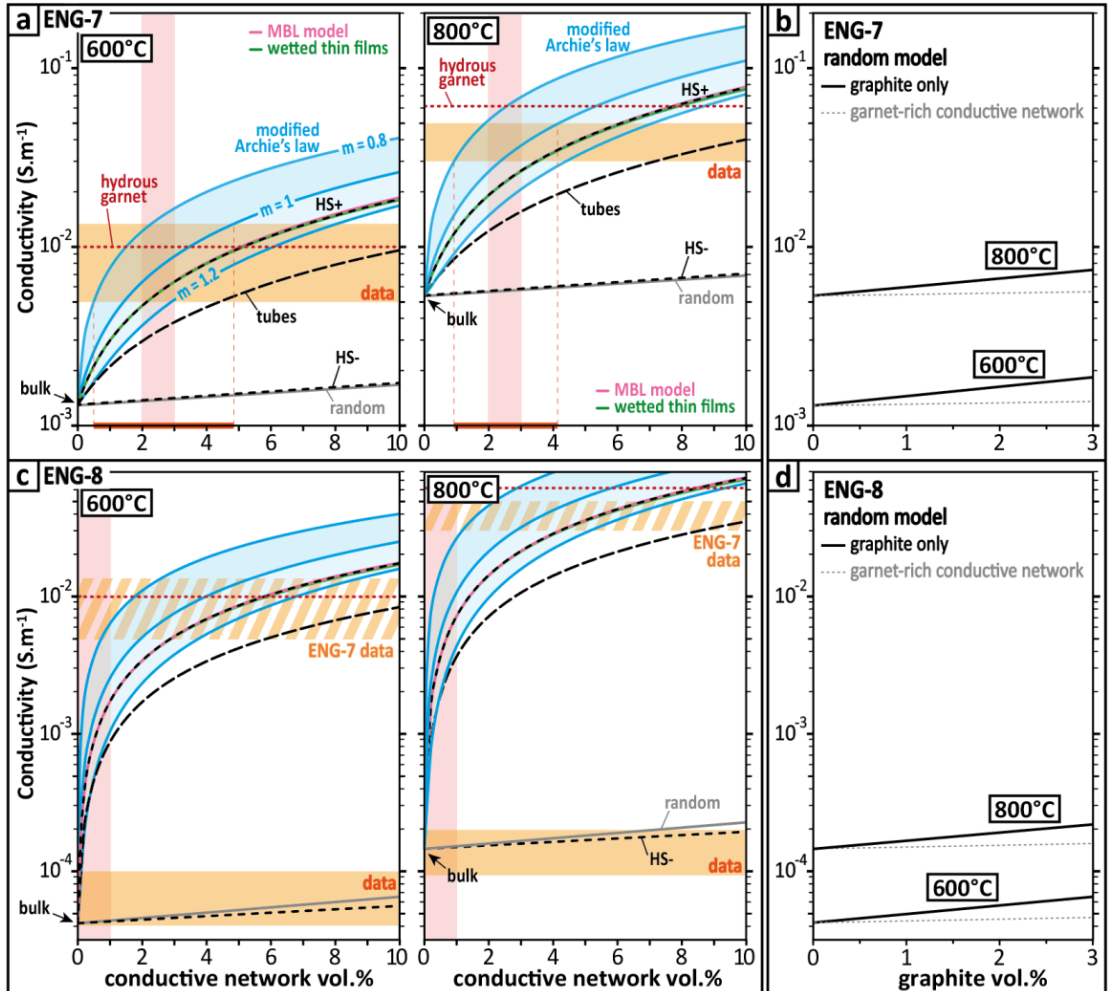

**Figure S12: Conductivity simulations considering various geometries.** Calculations for ENG7 (**a-b**) and ENG8 (**c-d**) at 600 and 800°C (no partial melting involved). The parameters considered for the models are listed in **Table S5**, and the formula are detailed in **Table S6**. Here the amount of $H_2O$ in garnet is assumed to be 465 ppm. See **Fig.15** for other amounts of $H_2O$. A table of values is provided in **Table S7**.

Table S5: Parameters used for electrical conductivity simulations.

| | Bulk silicate composition* | | | | | Carbon |
|---|---|---|---|---|---|---|
| | olivine | cpx | opx | garnet | others | Graphite |
| ENG7 | 2 vol.% | 87 vol.% | 11 vol.% | 0 vol.% | *trace* | ~ 1 vol.% |
| ENG8 | 95 vol.% | 5 vol.% | 0 vol.% | 0 vol.% | *trace* | |
| | Water content | | | | | |
| | olivine | cpx | opx** | garnet*** | | |
| | 20 ppm | 275 ppm | 200 ppm | 0 – 465 ppm | | |
| | Corresponding conductivity (S.m$^{-1}$) | | | | | |
| Temperature | olivine | cpx | opx** | garnet | | graphite**** |
| | | | | dry | 465 ppm | |
| 600°C | $3.5 \times 10^{-5}$ | $1.5 \times 10^{-3}$ | $8 \times 10^{-4}$ | $3 \times 10^{-5}$ | $1 \times 10^{-2}$ | $1 \times 10^5$ |
| 800°C | $1.2 \times 10^{-4}$ | $6 \times 10^{-3}$ | $4 \times 10^{-3}$ | $3 \times 10^{-4}$ | $6 \times 10^{-2}$ | |
| | | | | 46 ppm | 160 ppm | |
| | | | 600°C | $5 \times 10^{-4}$ | $2 \times 10^{-3}$ | |
| | | | 800°C | $3 \times 10^{-3}$ | $1 \times 10^{-2}$ | |
| reference | Yoshino et al. (2012) | Yang et al. (2011) | Dai & Karato (2009) | Dai et al. (2012) Dai & Karato (2009) | | Duba & Shankland (1982) |

*ideal reconstructed compositions of the starting material (Chin, 2018).
**values for 200 ppm H$_2$O and X$_{Fe}$ = 0.1 (Dai & Karato, 2009), close to the actual opx in the xenoliths (≈ 160 ppm H$_2$O, X$_{Fe}$ ≈ 0.16). Some more H$_2$O and less Fe would increase and decrease the conductivity, respectively.
***garnet is (sub)micrometric and forms during runs (3 GPa), and its H$_2$O content is unknown. In **Fig.2** and **Fig.S12**, hydrous garnet (Dai et al., 2012) is considered. **Fig.S13** focuses on the impact of water content.
****graphitic carbon is expected in mantle xenoliths, but its actual fraction is uncertain (∼ 0.1 wt.%).

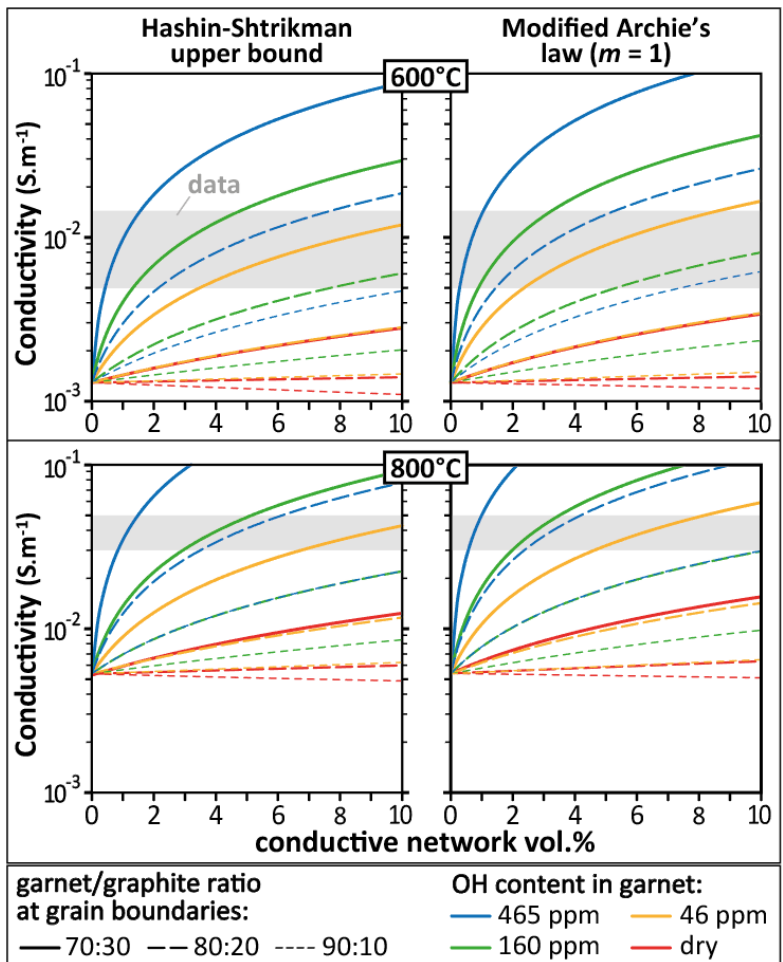

**Figure S13: Effect of garnet/graphite ratio and water content on conductivity.** Detailed simulations for the Hashin-Shtrikman upper bound (left) and the modified Archie's law (right) at 600 and 800°C. Simulation on ENG7 composition, compared to data (grey). The parameters considered for the models are listed in **Table S5**, and the formula are detailed in **Table S6.** A table of values is provided in **Table S7**.

**Table S6: various models to compute electrical conductivity in Fig.2, Fig.S12 and Fig.S13.**

| Model | Equation | Reference |
|---|---|---|
| Hashin-Shtrikman upper bound: | $\sigma^* = \sigma_n + \dfrac{(1-X_n)}{\dfrac{1}{\sigma_b - \sigma_n} + \dfrac{X_n}{3\sigma_n}}$ | Hashin & Shtrikman (1962) |
| Hashin-Shtrikman lower bound: | $\sigma^* = \sigma_b + \dfrac{(1-X_n)}{\dfrac{1}{\sigma_n - \sigma_b} + \dfrac{1-X_n}{3\sigma_n}}$ | Hashin & Shtrikman (1962) |
| Tubes model | $\sigma^* = \dfrac{1}{3} X_n \sigma_n + (1-X_n)\sigma_b$ | Grant & West (1965) |
| Wetted thin films model | $\sigma^* = \dfrac{\sigma_b \sigma_n (1-X_n)^{\frac{2}{3}}}{\sigma_n(1-X_n)^{\frac{1}{3}} + \sigma_b\left[1-(1-X_n)^{\frac{1}{3}}\right]} + \sigma_n\left[1-(1-X_n)^{\frac{2}{3}}\right]$ | Waff (1974) |
| Modified brick-layer model | $\sigma^* = \dfrac{\sigma_n\left[\sigma_n\left((1-X_n)^{\frac{2}{3}} - 1\right) - \sigma_n(1-X_n)^{\frac{2}{3}}\right]}{\sigma_b\left[(1-X_n) - (1-X_n)^{\frac{2}{3}}\right] + \sigma_n\left[(1-X_n)^{\frac{2}{3}} + X_n - 2\right]}$ | Partzsch (1997) |
| Modified Archie's law | $\sigma^* = \sigma_b(1-X_n)^p + \sigma_n X_n^m$ <br> $p = \dfrac{\log(1-X_n^m)}{\log(1-X_n)}$ | Glover et al. (2000) |

**Table S7: conductivity values (in S.m$^{-1}$) for different amounts and nature of conductive networks.** Calculations of ENG-7 composition using the Hashin-Shtrikman upper bound. As shown of **Fig.S12**, when the conductive network represents more than 1-2 vol.% of the sample, the ENG-7 and ENG-8 compositions display the same values, which means that the conductive network fully controls the conductivity. In addition, **Fig.S13** shows that these values are close to what is expected if we consider the modified Archie's law.

| conductive network = 2 vol.% | | 600°C | | | | 800°C | | | |
|---|---|---|---|---|---|---|---|---|---|
| | | OH content in garnet (ppm) | | | | OH content in garnet (ppm) | | | |
| | | 465 | 160 | 46 | 0 | 465 | 160 | 46 | 0 |
| garnet/ graphite ratio | 70:30 | 1.8.10$^{-2}$ | 6.8.10$^{-3}$ | 3.4.10$^{-3}$ | 1.6.10$^{-3}$ | 6.5.10$^{-2}$ | 2.2.10$^{-2}$ | 1.3.10$^{-2}$ | 5.6.10$^{-3}$ |
| | 80:20 | 4.7.10$^{-3}$ | 2.2.10$^{-3}$ | 1.6.10$^{-3}$ | 1.3.10$^{-3}$ | 1.9.10$^{-2}$ | 8.6.10$^{-3}$ | 6.5.10$^{-3}$ | 5.2.10$^{-3}$ |
| | 90:10 | 2.0.10$^{-3}$ | 1.5.10$^{-3}$ | 1.3.10$^{-3}$ | 1.3.10$^{-3}$ | 8.6.10$^{-3}$ | 5.9.10$^{-3}$ | 5.5.10$^{-3}$ | 4.7.10$^{-3}$ |
| | 100:0 | 1.4.10$^{-3}$ | 1.3.10$^{-3}$ | 1.3.10$^{-3}$ | 1.0.10$^{-3}$ | 6.1.10$^{-3}$ | 5.4.10$^{-3}$ | 5.3.10$^{-3}$ | 2.5.10$^{-3}$ |

| conductive network = 4 vol.% | | 600°C | | | | 800°C | | | |
|---|---|---|---|---|---|---|---|---|---|
| | | OH content in garnet (ppm) | | | | OH content in garnet (ppm) | | | |
| | | 465 | 160 | 46 | 0 | 465 | 160 | 46 | 0 |
| garnet/ graphite ratio | 70:30 | 3.5.10$^{-2}$ | 1.2.10$^{-2}$ | 5.5.10$^{-3}$ | 1.9.10$^{-3}$ | 1.2.10$^{-1}$ | 3.9.10$^{-2}$ | 2.0.10$^{-2}$ | 5.8.10$^{-3}$ |
| | 80:20 | 8.1.10$^{-3}$ | 3.2.10$^{-3}$ | 1.9.10$^{-3}$ | 1.3.10$^{-3}$ | 3.4.10$^{-2}$ | 1.2.10$^{-2}$ | 7.8.10$^{-3}$ | 5.2.10$^{-3}$ |
| | 90:10 | 2.6.10$^{-3}$ | 1.6.10$^{-3}$ | 1.4.10$^{-3}$ | 1.2.10$^{-3}$ | 1.2.10$^{-2}$ | 6.6.10$^{-3}$ | 5.7.10$^{-3}$ | 4.2.10$^{-3}$ |
| | 100:0 | 1.6.10$^{-3}$ | 1.3.10$^{-3}$ | 1.2.10$^{-3}$ | 8.2.10$^{-4}$ | 6.9.10$^{-3}$ | 5.5.10$^{-3}$ | 5.2.10$^{-3}$ | 1.6.10$^{-3}$ |

| conductive network = 6 vol.% | | 600°C | | | | 800°C | | | |
|---|---|---|---|---|---|---|---|---|---|
| | | OH content in garnet (ppm) | | | | OH content in garnet (ppm) | | | |
| | | 465 | 160 | 46 | 0 | 465 | 160 | 46 | 0 |
| garnet/ graphite ratio | 70:30 | 5.3.10$^{-2}$ | 1.8.10$^{-2}$ | 7.6.10$^{-3}$ | 2.2.10$^{-3}$ | 1.9.10$^{-1}$ | 5.7.10$^{-2}$ | 2.7.10$^{-2}$ | 6.1.10$^{-3}$ |
| | 80:20 | 1.2.10$^{-2}$ | 4.1.10$^{-3}$ | 2.2.10$^{-3}$ | 1.4.10$^{-3}$ | 4.8.10$^{-2}$ | 1.5.10$^{-2}$ | 9.1.10$^{-3}$ | 5.1.10$^{-3}$ |
| | 90:10 | 3.3.10$^{-3}$ | 1.8.10$^{-3}$ | 1.4.10$^{-3}$ | 1.2.10$^{-3}$ | 1.6.10$^{-2}$ | 7.2.10$^{-3}$ | 5.9.10$^{-3}$ | 3.8.10$^{-3}$ |
| | 100:0 | 1.7.10$^{-3}$ | 1.3.10$^{-3}$ | 1.2.10$^{-3}$ | 6.9.10$^{-4}$ | 7.6.10$^{-3}$ | 5.6.10$^{-3}$ | 5.1.10$^{-3}$ | 1.2.10$^{-3}$ |

# Supplementary text:

## S1 - Nucleation of hydrous garnet at grain boundaries during the sintering process

### a. Garnet identification

Garnet is identified for every run at 3 GPa. It exhibits an atoll shape (e.g. **Fig.S8**), typical for metamorphic garnet. Although the grain size is too small to get the exact formula using EDS measurements, it is sufficient to deduce the approximate formula ∼$Py_{51}$-$Alm_{18}$-$Grs_{31}$. In peridotites and pyroxenites, only two other minerals could have the same $(Mg,Ca,Fe)^{2+}/(Al,Fe)^{3+}/Si$ ratio as garnet (equal to 3/2/3):

- Mg-pumpellyite, which is a hydrous mineral (general formula $Mg_{6-x}Al_{4+x}Si_6O_{20+x}(OH)_{8-x}$ thus $Mg_3Al_2Si_3O_{10}(OH)_4$ for x =0) dehydrating at 600°C at a pressure of 3 GPa (Fockenberg, 1998), and which should have much more oxygen that what we can see in the EDS data. Pumpellyites are monoclinic sorosilicates, while garnet is a cubic or tetragonal nesosilicate.
- Al-rich orthopyroxene (solid solution $Mg_2Si_2O_6$-$MgAl_2SiO_6$, which contains a pyrope-like composition), but the Al content of such orthopyroxene is usually associated with low pressures only. Such Al-rich orthopyroxene cannot include Ca, which is inconsistent with the EDS data, revealing no Ca in orthopyroxene but substantial Ca in garnet (**Fig.S10** and **S11**).

As detailed above, both pumpellyite and Al-rich orthopyroxene are unlikely to explain the observations, in contrast with garnet.

### b. Garnet nucleation and growth

Reaction kinetics is key in the development of metarmorphic microtextures (Ridley & Thompson, 1986). Observations of garnet nucleation and growth at low temperatures are rare in the literature, but some studies have reported it (Lanari & Engi, 2017 and references therein). Notably, garnet crystals (5 mm average diameter) are reported in the southern Omineca belt of the Canadian Cordillera, which underwent Barrovian metamorphism peaking at middle amphibolite facies during the Early Cretaceous. Garnet porphyroblasts grew along a P-T path from 500°C at 0.5 GPa to 570°C at 0.7 GPa. For comparison, in experiment BB-253 (≤ 750°C, 3 GPa), submicrometric garnet crystals grew at grain boundaries in less than 2 hours at T > 400°C.

Most experimental petrology experiments focus on high temperatures for two reasons: 1) in nature, garnet nucleation and growth at T < 500°C in ultramafic rocks is rather unrealistic, and 2) high-temperatures are required for grain growth to occur fast enough during an experiment in order to perform single-grain analysis with various techniques.

Polymorphic transformations such as the quartz-coesite transition can easily occur at low temperatures because it requires no or limited atomic diffusion. Contrastingly, the "spinel-garnet transition", i.e. transition from a spinel lherzolite/pyroxenite to a garnet lherzolite/pyroxenite, is a metamorphic reaction involving the reorganization of several phases within a paragenesis, i.e. at the scale of several grains (e.g. Lanari & Engi, 2017, and references therein). For such metamorphic reaction, the thermodynamics is highly impacted by kinetics, which allows at low temperatures the preservation of metastable HP-HT rocks eventually exposed to the surface after the erosion of internal mountain belts. Especially, any reaction that involves garnet formation involves $Al^{3+}$ diffusion, which is slow at 800°C.

Another parameter controlling reaction kinetics is thermodynamic overstepping, which is a driving force that is necessary for porphyroblast nucleation and growth (Spear et al., 2017). In the present study, the growth of nanometric garnets initiated during the initial sintering process (**Methods, section M2**). At such low temperatures, the reaction occurred only where the reactive minerals are close to each other, i.e. at grain boundaries. Although $Al^{3+}$ intracrystalline diffusion is slow, I consider that the original garnet-rich network in the samples at temperatures between 400 and 800°C is the result of the small grain size and favoured contacts between the reactive minerals during initial heating and powder compaction.

Considering a Stefan's interface model for garnet growth, the growth of crystals at 700°C up to a radius of 10 or 100 nm is feasible in a few minutes and a few hours, respectively (Spear et al., 2017; Mg-poor system). Some garnet grains are larger (up to ≈ 1 µm in radius at 750°C; **Fig.S11**), which can be due to both the large pressure overstep (ΔP > 1 GPa; **Fig.S4**) and larger diffusivity in Mg-rich systems compared to Mg-poor systems.

### c. Widespread hydrous garnets in the mantle: hydrogarnet and other hydrous defects

Garnets can incorporate $H_2O$, via hydrogarnet substitution (substitution of $Si^{4+}$ by 4 $H^+$; **Fig.3d** Ackermann et al., 1983; Aines & Rossman, 1984; Wright et al., 1994; Katayama et al., 2003), or via a coupled substitution of $Si^{4+}$ by [$H^+ + Al^{3+}$] (Mookherjee & Karato, 2010). The hydrogarnet defect is stabilized by hydrogen bonding (**Fig.3d**; Lacivita et al., 2015). Hydrogarnet is generally more stable in grossular ($Ca_3Al_2Si_3O_{12}$ or $Grs_{100}$), as documented by several studies (Withers et al., 1998; Geiger & Rossman, 2018; 2020). However, it is demonstrably stable in pyrope ($Mg_3Al_2Si_3O_{12}$ or $Py_{100}$) at conditions relevant to the Tanzanian mantle (2.5 GPa, 1000°C; Ackermann et al., 1983). In the $MgO-Al_2O_3-SiO_2$ system, $H_2O$ contents in synthetic pyrope are ≈ 0.05 wt.% (500 ppm) and consist of $(HO)_4^{4-}$ clusters. The $H_2O$ content is ≈ 0.05 wt.% (500 ppm) in the Wesselton kimberlite (South Africa), 0.09 wt.% (900 ppm) beneath the Udachnaya Craton (Maldener et al., 2003) and 0.25 wt.% (2500 ppm) beneath the Colorado Plateau (Aines & Rossman, 1984). The garnet identified in this study (3 GPa) has the approximate formula ∼$Py_{51}$-$Alm_{18}$-$Grs_{31}$, which is, considering the explanations hereabove, ideal for hydrogarnet stability.

The only studies that have investigated the electrical conductivity of hydrous garnets used a pyrope from Garnet Ridge, Arizona, USA, of composition ∼$Py_{73}$-$Alm_{14}$-$Grs_{13}$ (Dai & Karato, 2009; Dai et al., 2012), with water contents of 465, 160 and 46 ppm. A value of 465 ppm is assumed in **Fig.2** and **Fig.S12**. The impact of OH content on the simulations is addressed in **Fig.S13**.

### d. Connectivity of garnet-rich networks

The garnet-rich network developing at grain boundaries corresponds to an example of a local mineral assemblage due to slow metamorphic reactions within the intergranular system, referred to as grain-boundary equilibrium (Lanari & Engi, 2017). Even though most garnet grains are much smaller at 750°C (BB-253; **Fig.S6** and **S11**) than at 1400°C (BB-246; **Fig.S7** and **S10**), the garnet network is already well developed, which explains the high conductivity even at low temperatures (500-750°C; **Fig.2**).

Garnet connectivity was reported in various natural rocks (Henjes-Kunst & Altherr, 1992; John et al., 2004; Evans et al., 2011; Smit et al., 2011; Vrijmoed et al., 2013; Lanari & Engi, 2017). The garnet-rich network developing at grain boundaries corresponds to an example of local mineral assemblage due to slow metamorphic reactions within the intergranular system, referred to as grain-boundary equilibrium (Lanari & Engi, 2017).

Garnet networks have been documented in highly strained hydrous eclogite mylonites from the Norwegian Caledonides (Smit et al., 2011), and at the sub-meter scale at least, in an eclogite from the Zambian fossil subduction zone, where grain-boundary sliding induced phase aggregation and intergranular pressure solution connected the Fe-rich inner cores of the garnets (Smit et al., 2011). Garnetite veins have also been described in peridotites metasomatized during the Caledonian orogeny (Vrijmoed et al., 2013). They allow a high garnet connectivity at the outcrop scale and the matrix of the metasomatized peridotites exhibits a poikilitic texture, which forms connected garnet network in between the garnetite veins (Vrijmoed et al., 2013). The mantle outcrops that show garnet connectivity (Vrijmoed et al., 2013) likely endured deformation and/or transformation on their way to the surface. Mantle xenoliths are the only direct access to the deep lithospheric mantle. Garnet-bearing mantle xenoliths are documented in the area, notably showing garnetite networks between large that grew at grain boundaries between large pyroxene grains (Henjes-Kunst & Altherr, 1992).

## S2 - Type and amount of carbon present & oxygen fugacity

Northern Tanzania is volcanically active – a prime example is Mt. Kilimanjaro – and the lithospheric mantle of the region is known to be $CO_2$-rich, as revealed by various xenoliths containing either carbonates (Lee & Rudnick, 1999; Lee et al., 2000) or carbonatite metasomatism (Rudnick et al., 1993). Xenoliths of the Lashaine area (10 km to the west from Engorora), contain about 3 wt.%. of $H_2O$ and 1.4 wt.% of $CO_2$ (Dawson et al., 1970), which suggests that the lithospheric mantle contains 10-100 ppm of $CO_2$ on average. Carbon could be easily identified and speciated using Raman spectrometry (e.g. Pearson et al., 1994). In the mantle, reductive conditions should favor graphite impurities at grain boundaries.

As illustrated in **Fig.S10** (BB-246, ENG7 pyroxenite), large garnet grains occur in clusters and contain significant chromium (Cr) and titanium (Ti), with minor manganese (Mn) and vanadium (V), while the garnet network (well highlighted by the Al map) is characterized by a simple and relatively homogeneous composition (Al, Ca, Fe). This compositional difference is explained by local heterogeneity, i.e. large garnet grains grow around grains containing higher contents of Cr, Ti and other heavy metals, which preferentially integrate the garnet structure in case of garnet nucleation. Some other minerals may contribute to increase conductivity, such as rutile ($TiO_2$) or chromite ($Cr_2O_3$), but only very locally. Minor phosphides seem to also take part of the metal-rich inclusions, as well as rare grains of metal alloys (**Fig.S10)**.

The preservation of the rhenium furnace, along with the stability of minor grains of Th-Ta-Ni alloy and the presence of titanium phosphide (**Fig.S10**) indicate a reducing environment, as expected in the cratonic mantle at 3 GPa (log $fO_2$ = -2; $\Delta FMQ$; Frost & McCammon, 2008). In such reduced conditions, carbon is more likely in the form of graphite that immobilizes at grain boundaries (Watson, 1986), consistently with the Raman spectra (**Fig.S13**). Conductive grain-boundary impurities can significantly impact the bulk conductivity, and their influence depends on their amount and nature (Watson et al., 2010).

Both the nature and repartition of the carbon impurities should be considered to evaluate the impact of carbon on electrical conductivity (Wang e t al., 2013). A previous study determined a percolation threshold of graphite GB impurities of $\sim$ 1 wt.% (Wang et al., 2013), while lower graphite contents ($\sim$ 0.1 wt.%) are associated with relatively low bulk conductivities (Watson et al., 2010). In this study, I assume the graphite content $\sim$ 0.1 wt.% (***Methods, section M6***), consistently with regional constraints (Dawson et al., 1970; Chin, 2018; **Table S1**), and evidence the nucleation of nanogarnet at grain boundaries allowing the involvement of the graphite impurities in the conductive network (**Fig.S12**, **Fig.S13**).

In addition, carbon solubility in olivine is relatively low compared to pyroxenes and garnet, but increases with pressure (Keppler et al., 2003). As ENG8 consists of 95% olivine, it is possible that the observed drop in conductivity with increasing pressure is due to the effect of pressure as well as enhanced carbon solubility.

## S3 - Remarks on the electromagnetic (EM) profile

Our results highlight that high-conductivity anomalies in the deep Tanzanian lithosphere could well be explained by garnet-rich networks (see main text), assisted by accessory graphitic impurities. In contrast, shallow anomalies ($\leq$ 35 km depth; $\approx$ 1.1 GPa) require other explanations, such as magma networks. Such magma networks may correspond to the mature rift system of the Central Vallee, Ethiopia, where near-surface high conductivity ($\sim$1 S.m$^{-1}$) is demonstrably related to magmatic activity (Keir et al., 2009). This is also the case for the Eyasi rift and Kilimanjaro regions at similar depths (**Fig.4**), where connected networks of volatile-bearing basaltic melts well explain (**Fig.2b**) shallow high-conductivity anomalies ($\geq 10^{-1}$ S.m$^{-1}$; $\leq$ 1 GPa; Selway, 2015).

The EM profile by Selway (2015) reinterpreted in this study in light of our experimental results has limitations inherent to either the geological context or the geophysical method. Because the entire lower portions of the lithospheric mantle of the Tanzanian craton appear as a uniform conductor in the inversion model presented by Selway (2015), it is not possible to identify the lower electrical bound of the conductivity anomaly. If graphite acts as an important connector of the garnet grains, I expect this mechanism to switch off upon the transition from the graphite to the diamond stability zone at a depth of ∼ 150 km, or most certainly at a depth above the LAB (**Fig.4e**).

## Supplementary references:

If not hereafter, studies cited in the SI are listed in the primary reference list.

1. Archie, G. E. 1942. The electrical resistivity log as an aid in determining some reservoir characteristics. *Transactions of the AIME* **146**(01), 54-62.
2. Clerc, J. P., Giraud, G., Alexander, S., Guyon, E. 1979. Conductivity of a mixture of conducting and insulating grains: Dimensionality effects. *Physical Review B.* **22** (5): 2489-2494.
3. Dai, L. & Karato, S. I. 2009. Electrical conductivity of orthopyroxene: Implications for the water content of the asthenosphere. *Proceedings of the Japan Academy, Series B* **85**(10), 466-475.
4. Dawson, J. B., Powell, D. G. & Reid, A. M. 1970. Ultrabasic xenoliths and lava from the Lashaine volcano, northern Tanzania. *Journal of Petrology* **11**(3), 519-548.
5. Fockenberg, T., 1998. An experimental study of the pressure-temperature stability of MgMgAl-pumpellyite in the system MgO-$Al_2O_3$-$SiO_2$-$H_2O$. *American Mineralogist* **83** (3-4), 220-227.
6. Foley, S. F. & Fischer, T. P. 2017. An essential role for continental rifts and lithosphere in the deep carbon cycle. *Nature Geoscience*, **10**(12), 897-902.
7. Frost, D. J. & McCammon, C. A. 2008. The redox state of Earth's mantle. *Annual Review of Earth & Planetary Sciences* **36**, 389-420.
8. Geiger, C. A. & Rossman, G. R. 2020. Micro- and nano-size hydrogarnet clusters and proton ordering in calcium silicate garnet: part I. The quest to understand the nature of "water" in garnet continues. *American Mineralogist*.
9. Geiger, C. A. & Rossman, G. R. 2018. IR spectroscopy and OH-in silicate garnet: the long quest to document the hydrogarnet substitution. *American Mineralogist* **103**(3), 384-393.
10. Glover, P. W., Hole, M. J. & Pous, J. 2000. A modified Archie's law for two conducting phases. *Earth & Planetary Science Letters* **180**(3-4), 369-383.
11. Grant, F. S. & West, G. F. 1965. *Interpretation theory in applied geophysics*. McGraw-Hill Book.
12. Green, D. H. & Hibberson, W. 1970. The instability of plagioclase in peridotite at high pressure. *Lithos* **3**(3), 209-221.
13. Gudfinnsson, G. H. & Presnall, D. C. 1996. Melting relations of model lherzolite in the system CaO-MgO-$Al_2O_3$-$SiO_2$ at 2.4-3.4 GPa and the generation of komatiites. *Journal of Geophysical Research: Solid Earth* **101**(B12), 27701-27709.
14. Gueguen, Y. & Dienes, J. 1989. Transport properties of rocks from statistics and percolation. *Mathematical geology* **21**(1), 1-13.
15. Hashin, Z. & Shtrikman, S. 1962. A variational approach to the theory of the effective magnetic permeability of multiphase materials. *Journal of applied Physics* **33**(10), 3125-3131.
16. Katayama, I., Hirose, K., Yurimoto, H. & Nakashima, S. 2003. Water solubility in majoritic garnet in subducting oceanic crust. *Geophysical Research Letters* **30**(22).
17. Keppler, H., Wiedenbeck, M. & Shcheka, S. S. 2003. Carbon solubility in olivine and the mode of carbon storage in the Earth's mantle. *Nature* **424**(6947), 414-416.
18. Lacivita, V., Mahmoud, A., Erba, A., D'Arco, P. & Mustapha, S. 2015. Hydrogrossular, $Ca_3Al_2(SiO_4)_{3-x}(H_4O_4)_x$: An ab initio investigation of its structural and energetic properties. *American Mineralogist* **100**(11-12), 2637-2649.
19. Lanari, P. & Engi, M. 2017. Local bulk composition effects on metamorphic mineral assemblages. *Reviews in Mineralogy & Geochemistry* **83**(1), 55-102.
20. Lee, C. T., Rudnick, R. L., McDonough, W. F. & Horn, I. 2000. Petrologic and geochemical investigation of carbonates in peridotite xenoliths from northeastern Tanzania. *Contributions to Mineralogy & Petrology* **139**(4), 470-484.
21. Miller, K. J., Montési, L. G. & Zhu, W. L. 2015. Estimates of olivine-basaltic melt electrical conductivity using a digital rock physics approach. *Earth & Planetary Science Letters* **432**, 332-341.

22. Mookherjee, M. & Karato, S. I. 2010. Solubility of water in pyrope-rich garnet at high pressures and temperature. *Geophysical Research Letters* **37**(3).
23. Partzsch, G.M. 1997. *Elektrische Leitfähigkeit partiell geschmolzener Gesteine: Experimentelle Untersuchungen, Modellrechnungen und Interpretation einer elektrisch leitfähigen Zone in den zentralen Anden.* Ph.D. Thesis, Freie Universität, Berlin.
24. Pommier, A. & Leinenweber, K. D. 2018. Electrical cell assembly for reproducible conductivity experiments in the multi-anvil. *American Mineralogist* **103**(8), 1298-1305.
25. Ridley, J. & Thompson, A. 1986. The role of mineral kinetics in the development of metamorphic microtextures. *Advances in physical geochemistry* **5**, 154-193.
26. Rudnick, R. L., McDonough, W. F. & Chappell, B. W. 1993. Carbonatite metasomatism in the northern Tanzanian mantle: petrographic and geochemical characteristics. *Earth & Planetary Science Letters* **114**(4), 463-475.
27. Smit, M. A., Scherer, E. E., John, T. & Janssen, A. 2011. Creep of garnet in eclogite: Mechanisms and implications. *Earth & Planetary Science Letters* **311**(3-4), 411-419.
28. Spear, F. S. 2017. Garnet growth after overstepping. *Chemical Geology* **466**, 491-499.
29. Stauffer, D. & Aharony, A. 1992. *Introduction to Percolation Theory* 2nd ed. Taylor & Francis.
30. Waff, H. S. 1974. Theoretical considerations of electrical conductivity in a partially molten mantle and implications for geothermometry. *Journal of Geophysical Research* **79**(26), 4003-4010.
31. Watson, H. C., Roberts, J. J. & Tyburczy, J. A. 2010. Effect of conductive impurities on electrical conductivity in polycrystalline olivine. *Geophysical Research Letters* **37**(2).
32. Watson, E. B. 1986. Immobility of reduced carbon along grain boundaries in dunite. *Geophysical Research Letters* **13**(6), 529-532.
33. Withers, A. C., Wood, B. J. & Carroll, M. R. 1998. The OH content of pyrope at high pressure. *Chemical Geology* **147**(1-2), 161-171.
34. Wright, K., Freer, R. & Catlow, C. R. A. 1994. The energetics and structure of the hydrogarnet defect in grossular: a computer simulation study. *Physics & Chemistry of Minerals* **20**(7), 500-503.
35. Zhu, W., Gaetani, G. A., Fusseis, F., Montési, L. G. & De Carlo, F. 2011. Microtomography of partially molten rocks: three-dimensional melt distribution in mantle peridotite. *Science* **332**(6025), 88-91.